\documentclass[pdflatex,sn-mathphys-num]{sn-jnl}

\usepackage{graphicx}%
\usepackage{multirow}%
\usepackage{amsmath,amssymb,amsfonts}%
\usepackage{amsthm}%
\usepackage{mathrsfs}%
\usepackage[title]{appendix}%
\usepackage{xcolor}%
\usepackage{textcomp}%
\usepackage{manyfoot}%
\usepackage{booktabs}%
\usepackage{algorithm}%
\usepackage{algorithmicx}%
\usepackage{algpseudocode}%
\usepackage{listings}%
\usepackage{braket}
\usepackage{float}

\theoremstyle{thmstyleone}%
\theoremstyle{thmstyletwo}%

\theoremstyle{thmstylethree}%
\newtheorem{definition}{Definition}%

\theoremstyle{thmstyleone}%
\providecommand{\KL}{D_{\mathrm{KL}}}
\providecommand{\Mo}{\mathrm{Mo}}
\providecommand{\Co}{\mathrm{Co}}
\begin{document}

\title[Article Title]{QuantumXCT: Learning Interaction-Induced State Transformation in Cell-Cell Communication via Quantum Entanglement and Generative Modeling}


\author*[1,2,3]{\fnm{Selim} \sur{Romero}}\email{ssromerogon@tamu.edu}

\author[1,3]{\fnm{Shreyan} \sur{Gupta}}\email{xenon8778@tamu.edu}

\author[2,3]{\fnm{Robert S.} \sur{Chapkin}}\email{r-chapkin@tamu.edu}

\author*[1,3,4]{\fnm{James J.} \sur{Cai}}\email{jcai@uga.edu}

\affil[1]{\orgdiv{Department of Veterinary Integrative Biosciences}, \orgname{Texas A\&M University}} 

\affil[2]{\orgdiv{Department of Nutrition}, \orgname{Texas A\&M University}}

\affil[3]{\orgdiv{CPRIT Single Cell Data Science Core}, \orgname{Texas A\&M University}}

\affil[4]{\orgdiv{Department of Biochemistry and Molecular Biology}, \orgname{University of Georgia}}


\abstract{
Inferring cell-cell communication (CCC) from single-cell transcriptomic data remains challenging. Typically, most existing computational methods take an effective but largely static and indirect strategy. They retrieve curated ligand-receptor gene pairs, calculate interaction scores from coordinated average expression, and then apply statistical testing. These methods tend to overlook intracellular signaling consequences, context dependence involving multiple ligand-receptor pairs, and higher-order interactions. Here, we introduce QuantumXCT, a hybrid quantum-classical framework that defines CCC as learnable transformations between cellular state distributions induced by cell-cell interaction. By encoding transcriptomic profiles into a high-dimensional Hilbert space, QuantumXCT trains parameterized quantum circuits to learn a unitary transformation mapping a baseline non-interacting cellular state to an interacting state. CCC is inferred from interaction-driven changes in cellular state distributions under explicitly stated modeling assumptions. Unlike methods restricted to curated ligand-receptor pairs, QuantumXCT can evaluate any cross-cell gene pair. We validate QuantumXCT using synthetic data with known ground-truth interactions and real data derived from an ovarian cancer-fibroblast mono- and co-culture model. In addition, on synthetic data, QuantumXCT ranked the programmed intercellular interactions above non-interacting gene pairs (ROC-AUC 0.93). On the co-culture data, it identified a \texttt{PDGFB-PDGFRB-STAT3} axis reproducible across bootstrap resamples, amplitude encodings and search algorithms, and consistent with published perturbation experiments. Importantly, the learned quantum circuit is interpretable: its entangling topology was translated into biologically meaningful interaction networks, while post hoc contribution analysis quantified the relative influence of individual interactions on the observed state transitions. By shifting CCC inference from static interaction lookup toward learning data-driven state transformations that reflect context-specific communication programs across cellular states and conditions, QuantumXCT provides a framework for generative modeling of intercellular communication. This work provides an interpretable, hypothesis-generating framework for \textit{de novo} discovery of communication programs, and delineates both the potential and the current limits of quantum machine learning in single-cell biology.
}

%
%
%


\maketitle

\section{Introduction}\label{sec_intro}
Inferring cell-cell communication (CCC) from single-cell transcriptomic data is challenging but crucial for understanding complex biological systems. Over the past decade, a growing ecosystem of computational tools has emerged to address this problem, ranging from ligand-receptor scoring methods such as CellChat \cite{jin2021inference, jin2025cellchat} and CellPhoneDB \cite{efremova2020cellphonedb} to multi-method frameworks such as LIANA \cite{dimitrov2022comparison} and semi-supervised approaches such as scTenifoldXct \cite{yang2023sctenifoldxct}. Despite this diversity, most current methods remain tethered to curated databases of known ligand-receptor pairs \cite{armingol2024diversification}. This reliance creates a critical bottleneck: it constrains discovery to existing knowledge and limits the identification of novel or context-specific signaling pathways. A systematic comparison of sixteen inference resources and seven methods showed that both resource and method selection strongly influence predicted interactions, underscoring the instability of database-dependent approaches \cite{dimitrov2022comparison}. Moreover, as we have recently demonstrated in a simulation study  \cite{romero2026qsimcells}, standard CCC inference methods fail to recover programmed causal pathways and instead report spurious associations driven by expression statistics alone. At their core, existing methods are primarily designed to test co-expression of known interacting genes rather than learn the underlying principles of communication directly from the data \cite{armingol2024diversification}.

To overcome these limitations, we propose a paradigm shift in CCC inference. Rather than searching for interacting ligand-receptor pairs, we treat CCC inference as a problem modeling the emergent transformation of cellular states induced by interaction. Specifically, we hypothesize that the communication signal is quantitatively captured by the difference between two transcriptomic probability distributions: a baseline state of cells in isolation (e.g., mono-culture) and a perturbed state measured in an interacting context (e.g., co-culture). Defining this high-dimensional distributional shift, without imposing prior biological assumptions, requires a new class of computational modeling.

For the targeted gene panels analyzed here the transformation remains classically simulable, and we do not claim a computational quantum advantage. Our motivation for a quantum formulation is representational: the task maps naturally onto the strengths of quantum circuits. Quantum systems are inherently probabilistic and excel at representing the complex, high-dimensional probability landscapes that are difficult to capture with classical computers. A quantum processor with $n$ qubits can naturally explore a state space of $2^n$ dimensions, mirroring the possibly immense combinatorial complexity of a cell's transcriptome. Leveraging superposition and entanglement, a parameterized quantum circuit (PQC) \cite{benedetti2019parameterized} can be trained to approximate intricate probability distributions, making it a promising tool for capturing subtle, system-wide shifts induced by cell-cell interactions. This capability is harnessed within a hybrid quantum-classical pipeline \cite{cerezo2021variational} in which the circuit's structure and strength are learned in separate stages. A classical density-matrix screen first prunes the candidate entangling gates. Subsequently, a search over this reduced set selects the entangling topology, with a deliberate trade-off between cost and thoroughness. Three strategies span this trade-off: an exhaustive pairwise-permutation search (N-Wise, $n=2$) that is expensive but resists local minima; a cheaper multi-epoch search whose exploration is tunable through the number of epochs; and a forward-looking variational-quantum (VQE) routine that uses quantum hardware for the combinatorial gate selection, handing a well-curated candidate list to the inexpensive multi-epoch search that then orders it. A classical optimizer then refines the continuous gate angles to minimize a data-driven cost function.

Crucially, the trained PQC is not a black box; it provides an interpretable model of the underlying dynamics, under modeling assumptions that we state explicitly (Section~\ref{subsec:identifiability}). The optimized parameters of the learned unitary transformation encapsulate the rules governing cellular state changes, reflecting both intercellular signaling and downstream intracellular responses, such as gene regulatory network (GRN) rewiring. By learning the transformation itself, our quantum machine learning (QML) approach yields a data-driven representation of interactions. Moreover, by comparing circuits optimized under different biological conditions, we can systematically identify how communication programs are altered across development or disease.

\begin{figure}[h!]
    \centering
    \includegraphics[width=\textwidth]{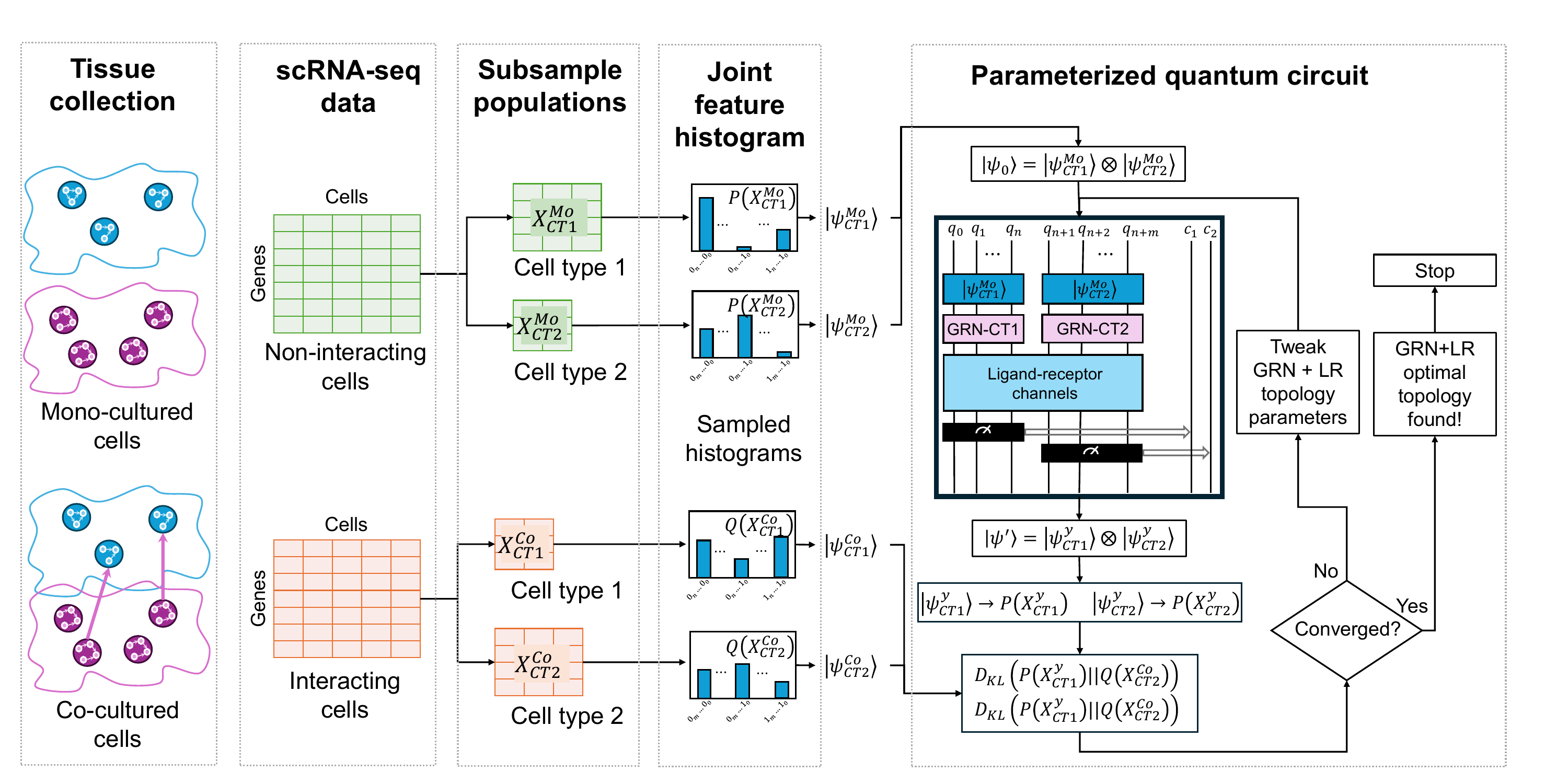}
    
    \caption{\textbf{A Quantum Generative Framework for Modeling Cell-Cell Communication.} 
    Our workflow reframes CCC as a learned transformation between cellular states. We derive two sets of empirical probability distributions from single-cell RNA sequencing: a \textbf{baseline distribution}, $P(X^{\text{Mo}})$, from non-interacting (mono-culture) cells, and a \textbf{target distribution}, $Q(X^{\text{Co}})$, from interacting (co-culture) cells. A PQC is then trained to model this transformation. The circuit is initialized to the global baseline state $\ket{\Psi_{\text{Mo}}}$ (derived from $P(X^{\text{Mo}})$) and applies a parameterized unitary operator, $U(\tau, \theta)$, to produce a final state whose simulated marginal distributions are denoted $P_{\psi'}$. In a hybrid quantum-classical loop, an optimizer iteratively adjusts the circuit's topology ($\tau$) and parameters ($\theta$) to minimize the KL divergence, $D_{\text{KL}}(P_{\psi'} \parallel Q_{\text{Co}})$, between the circuit's simulated output and the empirical target. Upon convergence, the optimized unitary, $U(\tau_{\text{opt}}, \theta_{\text{opt}})$, provides a quantitative, data-driven model of the communication dynamics.
    }
    \label{fig:scheme}
\end{figure}

To achieve this, we introduce a novel, database-free hybrid quantum-classical framework for learning CCC dynamics. Our approach centers on training a PQC to function as a \textit{generative transformation model}. Rather than adopting a standard classification-oriented Quantum Neural Network (QNN) architecture, we pioneer a method where the circuit learns to execute the unitary transformation that evolves a non-interacting cellular state into an interacting one. The optimization of the circuit's entangling topology and its gate parameters is guided by a custom-designed cost function based on the Kullback-Leibler (KL) divergence, which minimizes the difference between the circuit's output distribution and the empirically observed interacting cells' data (Fig.~\ref{fig:scheme}). We demonstrate this methodology on single-cell RNA sequencing (scRNA-seq) data, establishing a new avenue for the \textit{de novo} discovery of cellular communication channels. The Supplementary Information (SI) provides formal definitions, null models and additional benchmarks.

\section{Methods}\label{sec:methods}
Our methodology infers CCC by learning a quantum-mechanical transformation that models how cellular states shift from a non-interacting to an interacting environment. The framework comprises three main stages: (1) pre-processing single-cell transcriptomic data to define baseline and target distributions; (2) specify a PQC to represent the state transformation; and (3) use a multi-stage hybrid optimization strategy to learn circuit parameters that best capture the communication dynamics.

\subsection{Data Processing and Quantum State Encoding}
\label{subsec:processing_encoding}
Our framework takes as input scRNA-seq data from two distinct cell types (e.g., CT1 and CT2) collected under, both non-interacting (e.g., mono-culture) and interacting (e.g., co-culture) conditions (Fig.~\ref{fig:scheme}). The data were pre-processed using Scanpy \cite{wolf2018scanpy}, including total-count normalization and logarithmic transformation, $X' = \texttt{log}(1+X)$. After pre-processing, we selected, for each cell type, a subset of genes/features most relevant to the biological context under study. The qubit-to-gene mapping and register ordering are defined in SI Section~S1.

To map the continuous expression profile onto the discrete basis states of the quantum system, we binarized the expression of each selected gene. Each gene is considered ``active" (state 1) or ``inactive" (state 0) using a defined threshold; here, a gene was considered active if its log-normalized expression was greater than zero. This binarization was biologically principled since scRNA-seq data is inherently sparse, and zero observations often reflect genuinely low-expression states rather than purely technical dropout \cite{qiu2020embracing}. Binary representations of scRNA-seq data have been shown to faithfully capture cell-type-level biological variation \cite{bouland2021differential}, and the active/inactive threshold  typically mirrors the characteristic ON/OFF expression switching observed in regulatory genes at single-cell resolution \cite{korthauer2016statistical}. This procedure converts each cell's transcriptomic profile into a  bitstring, $s \in \{0, 1\}^d$, where $d$ is the number of selected genes. From the binarized data, we subsequently computed empirical joint frequency counts, $C(s)$ for each unique cellular state $s$.

These frequency counts are then used to construct the two primary inputs for our quantum model: the initial quantum state and the classical target distributions.
\begin{enumerate}
    \item \textbf{Baseline State Amplitudes:} To prepare the initial state of the quantum circuit, counts from the non-interacting (mono-culture) cells, $C_{\text{Mo}}(s)$, were L2-normalized to define the state vector's amplitudes, $\alpha^{\text{Mo}}_s$:
    \begin{equation}
        \alpha^{\text{Mo}}_s = \frac{C_{\text{Mo}}(s)}{\sqrt{\sum_{s' \in \{0,1\}^d} C_{\text{Mo}}(s')^2}}.
        \label{eq:amplitude_mo}
    \end{equation}
    These amplitudes define the individual \textbf{baseline state vectors}, $\ket{\psi^{\text{Mo}}_{\text{CT1}}}$ and $\ket{\psi^{\text{Mo}}_{\text{CT2}}}$.

    \item \textbf{Target Probability Distributions:} To define the classical target for our cost function, the counts from the interacting (co-culture) cells, $C_{\text{Co}}(s)$, were first L2-normalized to produce a vector of target amplitudes, $\alpha^{\text{Co}}_s$. These amplitudes were then squared to yield the final target probability distributions, $Q_{\text{Co}}(s)$:
    \begin{equation}
        \alpha^{\text{Co}}_s = \frac{C_{\text{Co}}(s)}{\sqrt{\sum_{s'} C_{\text{Co}}(s')^2}} \quad \text{and} \quad Q_{\text{Co}}(s) = |\alpha^{\text{Co}}_s|^2.
        \label{eq:prob_co}
    \end{equation}
\end{enumerate}
We consider two amplitude encodings. The L2 encoding of Eqs.~\eqref{eq:amplitude_mo}--\eqref{eq:prob_co} has measurement probabilities proportional to squared counts, $C(s)^2/\sum_{s'}C(s')^2$, which up-weight common states. The square-root (L1) encoding
\begin{equation}
\alpha_s=\sqrt{C(s)\big/\textstyle\sum_{s'}C(s')}
\label{eq:amplitude_l1}
\end{equation}
reproduces the empirical frequencies exactly, $|\alpha_s|^2=\hat p(s)$. The same encoding is always applied to the initial state, the target and the candidate screening. Both encodings are compared in Section~\ref{subsec:disc_encoding} and SI Section~S2, and L1 is the default of the released software.

\subsection{The Quantum Transformation Model}
\label{subsec:quantum_model}
Our model is a PQC \cite{benedetti2019parameterized} acting on a combined system of $N=n_1+n_2$ qubits, comprising an $n_1$-qubit logical register for CT1 and an $n_2$-qubit logical register for CT2. The circuit's evolution is described by the unitary transformation:
\begin{equation}
    \ket{\psi'} = U(\tau, \theta) \ket{\Psi_{\text{Mo}}}.
    \label{eq:unitary_evolution}
\end{equation}
Here, $\ket{\Psi_{\text{Mo}}}$ is the \textbf{global initial state}, representing the two cell types in isolation. It was constructed by taking the tensor product of the individual baseline state vectors (Eq.~\ref{eq:amplitude_mo}):
\begin{equation}
    \ket{\Psi_{\text{Mo}}} = \ket{\psi^{\text{Mo}}_{\text{CT1}}} \otimes \ket{\psi^{\text{Mo}}_{\text{CT2}}}.
    \label{eq:initial_state_global}
\end{equation}
The operator $U(\tau, \theta)$ models the intercellular communication and is composed of a sequence of entangling gates connecting the two logical registers. It is defined by its discrete \textbf{topology} $\tau$ and continuous \textbf{parameters} $\theta$.

\subsection{Hybrid Optimization Strategy}
\label{subsec:optimization}
The PQC was trained using a classical optimization loop \cite{cerezo2021variational} to find the optimal circuit parameters $(\tau_{\text{opt}}, \theta_{\text{opt}})$ that minimize a cost function based on the KL divergence \cite{kullback1951information}.

\subsubsection{Objective Function}
The cost function, $\mathcal{L}(\tau, \theta)$, measures the dissimilarity between the marginal distributions of the circuit's output state, $\ket{\psi'}$, and the target marginal distributions from the interacting data (Eqs.~\ref{eq:prob_co} and~\ref{eq:amplitude_l1}, for the L2 and L1 encodings respectively):
\begin{equation}
    \mathcal{L}(\tau, \theta) = D_{\text{KL}}(P_{\psi'}(\text{CT1}) \parallel Q_{\text{Co}}(\text{CT1})) + D_{\text{KL}}(P_{\psi'}(\text{CT2}) \parallel Q_{\text{Co}}(\text{CT2})),
    \label{eq:kl_cost}
\end{equation}
where $P_{\psi'}(\text{CT}_k)$ is the marginal probability distribution for cell type $k$ generated by the PQC as it evolves the non-interacting baseline state into the predicted interacting state. Note that the cost was computed over marginal distributions independently for each cell type, because mono- and co-culture experiments measure each population separately and no joint distribution over the two cell types exists. The model distribution is the first argument of $D_{\text{KL}}$ (a mode-seeking divergence). Probabilities are floored at $\varepsilon=10^{-9}$.

\subsubsection{Modeling Hypothesis: Interactions Carry the Change}\label{subsec:identifiability}
The mono-culture state already contains each gene's own basal activity (its ``self-rotation'') and the within-cell regulatory structure. QuantumXCT models the transition to co-culture as the interaction terms that are missing from this baseline. Every change must be produced by a controlled rotation whose control is a panel gene (one of the encoded genes), either in the partner cell type (candidate CCC) or in the same cell type (candidate GRN rewiring). Free single-qubit rotations are therefore deliberately excluded. This is the central hypothesis of the method, not a limitation to be relaxed for a better fit. Indeed, relaxation, by allowing free single-qubit rotations, would let separate rotations on each register match both marginals with no interaction at all (SI Section~S3). Two consequences follow from the structure of a controlled gate (SI Section~S3). First, a cross-cell gate $c\to t$ changes only the marginal of the target's cell type, so the \emph{responding} gene and cell type are identified. This direction reflects the marginal asymmetry of the controlled operation; the entanglement itself carries no direction (SI Section~S3). Second, the control gene acts through its basal activity, so controls with very similar activity profiles are nearly interchangeable. We therefore, read an edge $c\to t$ as ``the co-culture shift of gene $t$ is best explained as a response gated by the activity of gene $c$'', and test the inferred edges against null models, resampling and independent evidence (Section~\ref{sec:results}; SI Sections~S3, S4 and S8).

\subsubsection{Topology Search (\texorpdfstring{$\tau$}{tau})}
Finding the optimal entangling topology $\tau$ is a combinatorial sequence-search problem whose search space grows factorially with circuit length, and which we cast as a QUBO for Algorithm~3 \cite{lucas2014ising}, as it required selecting both the best subset of gates and their optimal application order. Because an exhaustive search over all gate subsets and orders is intractable, we first prune the candidate gates with a density-matrix heuristic and then search this reduced space with one of three algorithms (Algorithms 1--3 below).

For the discrete topology search, we used controlled-RX (CRX) gates with a fixed angle of $\pi/2$ as the basic building blocks. This choice enabled the detection of weaker correlations via partial entanglement, as detailed in the circuit construction pipeline in Appendix~\ref{appendix:circuit_construction}.

\paragraph{Heuristic Pruning of Candidate Gates} To guide the search, we first identified a high-potential candidate set of entangling gates, $\mathcal{C}$, by analyzing the difference between the global density matrices of the non-interacting and interacting systems (see Appendix~\ref{appendix:pruning} for details). By applying a threshold to the elements of the difference matrix, $\Delta\rho$ (Eq.~\ref{eq:dens_diff_appendix}), we generated a reduced list of candidate CRX($\pi/2$) gates that were most likely to facilitate the required state transformation. This pruned set $\mathcal{C}$ was the input to all subsequent search algorithms. For small panels the screen removes few candidates: for the 7-gene ovarian panel it retains 42, 28 and 12 of the 42 ordered pairs at $\delta\rho>0.005$, $0.01$ and $0.02$ (L1 encoding), so pruning mainly matters for larger panels (SI Section~S10).

\paragraph{Search Algorithms} 
We subsequently developed and compared three distinct strategies to search for the optimal topology within the pruned candidate set $\mathcal{C}$:
\begin{enumerate}
    \item \textbf{Algorithm 1 (Iterative Local Search):} A classical local-search heuristic \cite{hoos2018stochastic} that explored both gate selection and ordering through addition, insertion, and deletion moves, performing an exhaustive $n$-wise permutation search over candidate orderings at each addition step (pairwise at $n=2$). This makes it the most thorough but most computationally expensive of the three, and the most resistant to local minima (see Appendix~\ref{appendix:algo1} for full details).
    \item \textbf{Algorithm 2 (Multi-Epoch Sequential Construction):} A faster, stochastic greedy search that used multiple random starting points to build candidate sequences, followed by a refinement pruning step applying an Occam's Razor parsimony criterion \cite{blumer1987occam} (see Appendix~\ref{appendix:algo2} for full details).
    \item \textbf{Algorithm 3 (QUBO-based Variational Selection):} A quantum-native approach \cite{farhi2014quantum, peruzzo2014variational} that decoupled the problem into a gate-selection stage, solved with VQE or QAOA, and a final classical ordering (permutation) stage performed by the multi-epoch search (Algorithm~2) (see Appendix~\ref{appendix:algo3} for full details).
\end{enumerate}
Table~\ref{tab:algo_comparison} summarizes the strengths and weaknesses of each algorithm. Based on our empirical results, we selected Algorithm 2 for the primary analyses because it offers a strong balance between computational efficiency and solution quality. In practice, all three algorithms achieved similar levels of KL minimization. Algorithm 1 was the most stable but also the most computationally intensive. Algorithm 2 was faster and provides a reliable estimate of the achievable minimization, and Algorithm 3 was a forward-looking approach that used quantum-assisted gate selection to narrow down permutation candidates, with the potential to outperform Algorithm 1 given a sufficiently large initial candidate pool.

\begin{table}[htbp]
\centering
\caption{Comparison of Topology Search Algorithms}
\label{tab:algo_comparison}
\renewcommand{\arraystretch}{1.3}
\begin{tabular}{|p{0.2\textwidth}|p{0.35\textwidth}|p{0.35\textwidth}|}
\hline
\textbf{Algorithm} & \textbf{Strengths} & \textbf{Weaknesses} \\ \hline \hline
\textbf{1. Iterative Local Search} &
\begin{itemize}
    \item Holistically explores the combined selection and permutation space.
    \item Effective at finding complex, order-dependent interactions.
    \item Flexible search through addition, insertion, and deletion moves.
\end{itemize} &
\begin{itemize}
    \item The most computationally expensive of the three; cost grows with the candidate set and sequence length (exhaustive $n$-wise permutation at each addition step).
    \item A local-search heuristic, so no global-optimum guarantee, although it is the most resistant to local minima of the three.
\end{itemize} \\ \hline
\textbf{2. Multi-Epoch Sequential Construction} &
\begin{itemize}
    \item Computationally faster than full local search.
    \item Stochastic starting points reduce risk of getting stuck in poor local minima.
    \item Good balance of exploration and speed.
\end{itemize} &
\begin{itemize}
    \item Forward construction explores the permutation landscape less thoroughly than Algorithm~1.
    \item With a conservative number of epochs it can miss orderings in which a key gate must enter early; more epochs reduce this risk at added cost.
\end{itemize} \\ \hline
\textbf{3. QUBO-based Variational Selection} &
\begin{itemize}
    \item Maps the hard gate-selection problem to a physical ground state problem (path to quantum advantage).
    \item Can efficiently search over a vast number of gate combinations.
    \item Can be combined with classical methods for a hybrid approach.
\end{itemize} &
\begin{itemize}
    \item VQE/QAOA are heuristic: not guaranteed to reach the true ground state, though in practice they return a sufficiently good gate selection.
    \item Selects gate \emph{combinations} only; ordering is delegated to the multi-epoch search, so on its own it is weak at permutational exploration.
    \item That ordering stage inherits the combinatorial cost: for large selected sets it is as expensive as Algorithms~1--2, and the classical permutation step is unreliable beyond $k>8$--$10$ gates (Section~\ref{subsec:disc_marginal}).
\end{itemize} \\ \hline
\end{tabular}
\end{table}

\subsubsection{Parameter Optimization (\texorpdfstring{$\theta$}{theta})}
For finer-grained modeling, the fixed-angle CRX($\pi/2$) gates used in the discrete topology search are promoted to a continuous ansatz. This was achieved by replacing each fixed gate with a fully parameterized CRX($\theta_i$) gate, allowing the model to learn the precise interaction strength. The final optimization step solved for the optimal vector of continuous parameters:
\begin{equation}
    \theta_{\text{opt}} = \arg\min_{\theta} \mathcal{L}(\tau_{\text{opt}}, \theta),
\end{equation}
where $\theta$ is the angle set to be optimized. The gate angles are minimized with the gradient-free COBYLA \cite{powell1994direct}, with the topology $\tau_{\text{opt}}$ held fixed and starting from the topology-search value ($\theta_i=\pi/2$, the default of the released software), for both the shot-based and the exact (state-vector) objectives. Sensitivity to initialization, including optimization from $\theta_i=0$, is assessed in SI Section~S4 (Fig.~S2).

\subsection{Implementation Details}
All data processing and analyses were performed in Python. Single-cell data were handled using \texttt{scanpy} \cite{wolf2018scanpy}. The quantum circuits and simulations were implemented in IBM's Qiskit framework \cite{javadiabhari2024quantum, Qiskit}, using the \texttt{qiskit\_aer} simulator \cite{qiskit_aer}. Classical optimization used routines from \texttt{SciPy.optimize} \cite{virtanen2020scipy}. Circuits were simulated with 5000 seeded shots; contribution analyses were additionally evaluated from exact state-vector probabilities to remove shot noise (SI Section~S4). Post-optimization hardware validation on the IBM device \texttt{ibm\_kingston} was run through Qiskit Runtime and is reported in SI Section~S11. Scripts and tests reproducing all analyses are available in the public code repository. Code for our quantum generative framework is available at GitHub (\url{https://github.com/cailab-tamu/QuantumXCT}).

\section{Results}\label{sec:results}
We evaluated the performance of QuantumXCT in two distinct stages. First, we used a controlled synthetic dataset as a proof-of-concept to show that the framework recovers the intercellular interaction structure and the direction of the dominant links from the shift between the non-interacting and interacting states (details in Appendix~\ref{appendix:spatial_sim_cells}). Second, we applied the model to real scRNA-seq datasets to assess its ability to identify novel CCC channels in complex biological environments.

\subsection{Benchmarking via Synthetic State-to-State Transformations}\label{subsec:sim_results}
To establish the utility of our hybrid quantum-classical approach, we first benchmarked its ability to learn a high-dimensional probability distribution transformation using simulated cellular data. This synthetic dataset (Appendix~\ref{appendix:spatial_sim_cells}) provided a ground-truth environment, in which gene-gene and cell-cell interactions were explicitly defined, enabling quantitative evaluation of the accuracy of our topology search algorithms.

\subsubsection{Data Encoding and Experimental Setup}
Following the methodology in Section~\ref{subsec:processing_encoding}, we binarized the transcriptomic counts of selected gene sets into ordered bit-strings, where each bit represents the activity of a gene. These bit-strings define unique cellular states, which were then encoded as the basis states of a quantum system. The normalized frequency of these states determined the amplitudes of a state vector, $\ket{\psi}$, representing the entire cellular population in a high-dimensional Hilbert space. Throughout, the amplitude is the square root of each state's normalized frequency (the default L1 encoding); the L2 (escort) alternative is discussed in Section~\ref{subsec:disc_encoding}.

For this controlled benchmark, we defined CT1 using a 2-gene feature set ($G_{\text{CT1}} = \{g_{50}, g_{90}\}$) and CT2 using a 4-gene feature set ($G_{\text{CT2}} = \{g_{60}, g_{70}, g_{71}, g_{80}\}$). This configuration resulted in a 6-qubit system, where the first two qubits represent CT1 and the remaining four represent CT2.

\begin{figure}[H]
\centering
\includegraphics[width=\textwidth]{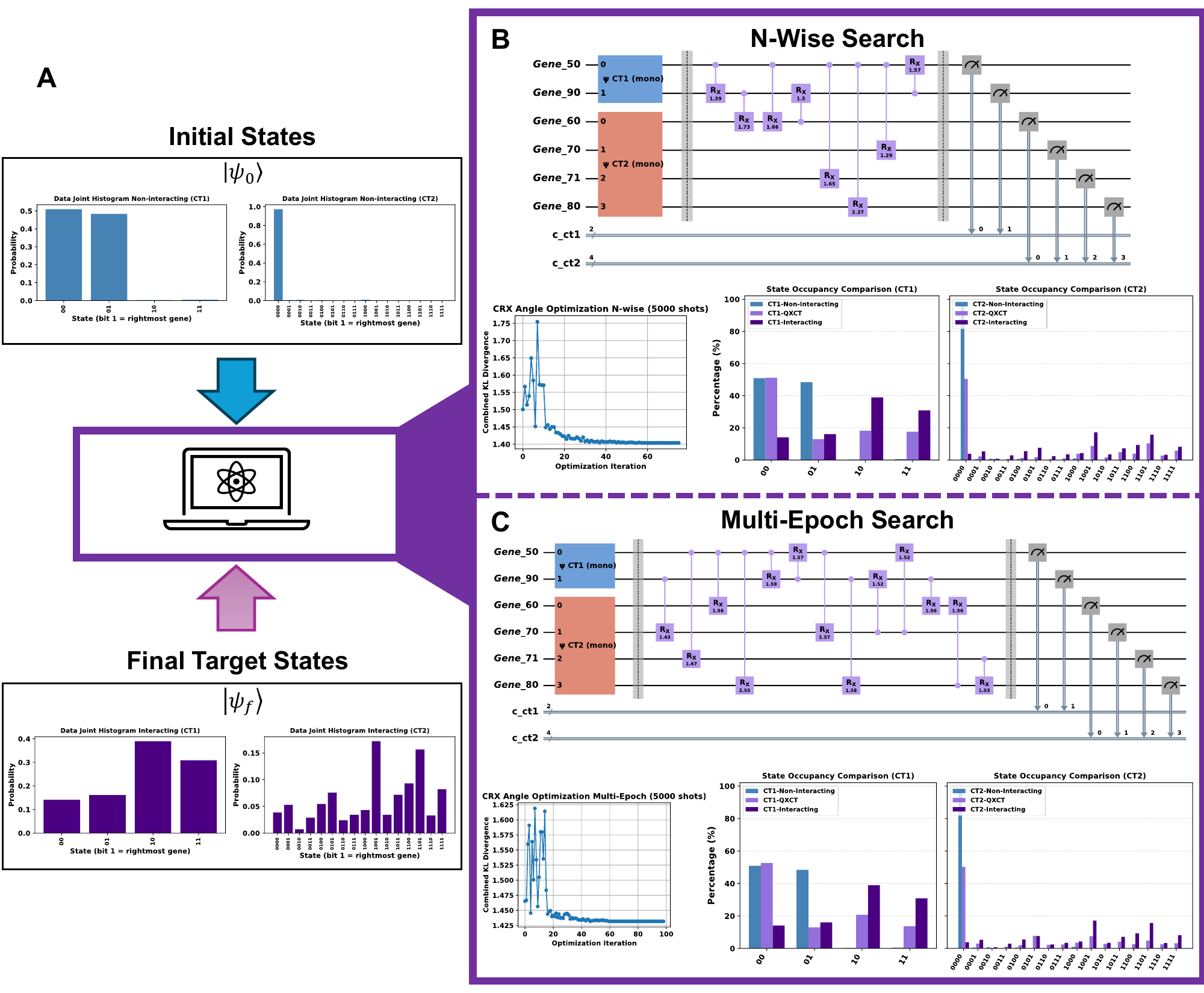}
\caption{\textbf{Validation of QuantumXCT on synthetic single-cell data (L1 encoding).} The framework learns the transformation from a non-interacting to an interacting cellular state on simulated data with ground-truth rules.
\textbf{(A) Optimization Workflow.} Input joint histograms of non-interacting cells define the \textbf{Initial States} ($\ket{\Psi_0}$), while interacting cell histograms define the \textbf{Final Target States} ($\ket{\Psi_f}$). The objective is to identify a unitary transformation, $U(\tau_{\text{opt}}, \theta_{\text{opt}})$, that evolves the initial distribution to best match the target.
\textbf{(B) N-Wise Search Performance.} Results for the Iterative Local Search (Algorithm 1, $n=2$). The panel shows the discovered quantum circuit topology, the KL divergence convergence during angle optimization, and the final state occupancy. The optimized QXCT distribution (light purple) is compared with the interacting ground-truth (dark purple).
\textbf{(C) Multi-Epoch Search Performance.} Corresponding results for the Multi-Epoch Search (Algorithm 2). Despite identifying a topologically distinct entangling circuit, the model achieves comparable convergence and occupancy matching. The state-occupancy and KL-convergence panels are computed from the 5000-shot Aer pipeline (the sampled estimator described in Methods); the per-gate contribution values quoted in the text and the network edge widths use exact state-vector probabilities (SI Section~S4).}
\label{fig:sim_comparison}
\end{figure}

As illustrated in Fig.~\ref{fig:sim_comparison}A, the joint histograms of non-interacting cells (``Initial States'') were used to construct the initial state vector, $\ket{\Psi_0}$, while the histograms from interacting cells (``Final Target States'') defined the target distributions for the KL divergence cost function. We deployed two of our primary classical search heuristics, the N-Wise Search (Algorithm~\ref{alg:local_search}, with $n=2$) and the Multi-Epoch Search (Algorithm~\ref{alg:multi_epoch}), to discover the optimal entangling topology. Both algorithms were used to identify the optimal unitary transformation, $U(\tau_{\text{opt}}, \theta_{\text{opt}})$, that evolves the initial state distribution to best match the target distribution, thereby learning the underlying communication rules from the data. Crucially, both heuristics converged to solutions of similar quality (76.3\% and 74.3\% loss reduction), despite discovering distinct circuit topologies (Fig.~\ref{fig:sim_comparison}B,~C).

\subsubsection{Fidelity of Learned Distributions and Topological Interpretability}
The optimization results for the synthetic benchmark demonstrate that QuantumXCT successfully learned the underlying causal structure of the cellular system. As shown in Fig.~\ref{fig:sim_comparison}, both the N-Wise and Multi-Epoch search heuristics converged to low KL divergence minima. While not a perfect replication, the output distributions of the optimized circuits (labeled ``QXCT") correctly modeled the key transformations from the non-interacting baseline. Both algorithms learned to suppress dominant non-interacting states while increasing the populations of interaction-induced states, demonstrating the model's ability to capture the directional shift in the high-dimensional probability landscape.

To validate the biological relevance of the discovered circuits, we mapped the entangling gates to the gene-qubit assignments: $\{q_0:g_{50}, q_1:g_{90}\}$ for CT1, and $\{q_2:g_{60}, q_3:g_{70}, q_4:g_{71}, q_5:g_{80}\}$ for CT2.

In the \textbf{N-Wise Search} (Fig.~\ref{fig:sim_comparison}B), the algorithm discovered a set of entanglements that represent a compressed, functionally equivalent model of the ground-truth pathways. It recovered the primary ligand-receptor interaction by entangling $q_0$ with $q_2$ ($g_{50} \to g_{60}$). Instead of learning the entire multi-step cascade literally, the algorithm utilized powerful ``functional shortcuts." For example, it directly entangled the initial driver $q_0$ ($g_{50}$) with the downstream gene $q_4$ ($g_{71}$), effectively summarizing the initial part of the CT2 GRN defined in Appendix~\ref{appendix:lr_grn_benchmark}. Similarly, it captured the full intercellular signaling arc by linking CT2's receptor $q_2$ ($g_{60}$) directly to CT1's final target $q_1$ ($g_{90}$). The search also placed reciprocal, both-direction gates between several node pairs ($g_{50}\leftrightarrow g_{90}$, $g_{50}\leftrightarrow g_{71}$ and $g_{60}\leftrightarrow g_{90}$), including back-edges into $q_0$ ($g_{50}$). The simulator contains no edge into $g_{50}$, and the entanglement itself carries no intrinsic direction (SI Section~S3): a reciprocal pair reflects the symmetric, mutual correlation that an entangling gate creates between two qubits, not directed feedback. They illustrate that the PQC learns efficient operations that reproduce the co-dependencies among key nodes ($g_{50}, g_{80}, g_{90}$), rather than recapitulating the wiring diagram edge by edge.

The \textbf{Multi-Epoch Search} (Fig.~\ref{fig:sim_comparison}C) arrived at a topologically distinct solution that nonetheless achieved comparable fidelity and reinforced the same biological principles. This heuristic identified $g_{50}$ as a central ``hub" by entangling its corresponding qubit $q_0$ ($g_{50}$) with multiple downstream genes in the CT2 cascade, $q_2$ ($g_{60}$), $q_4$ ($g_{71}$) and $q_5$ ($g_{80}$). Crucially, despite the different topology, it converged on the same key functional relationships found by the N-Wise search, such as links from the CT1 drivers $q_0$ ($g_{50}$) and $q_1$ ($g_{90}$) to the CT2 receptor $q_2$ ($g_{60}$). As in the N-Wise search, several pairs are entangled in both directions ($g_{60}\leftrightarrow g_{90}$, $g_{50}\leftrightarrow g_{90}$, $g_{50}\leftrightarrow g_{70}$ and $g_{80}\leftrightarrow g_{90}$), reflecting the symmetric correlation of the entangling gate rather than directed feedback. The emergence of these consistent entangling patterns across different search algorithms suggests that our framework is uncovering the dominant correlations dictated by the underlying GRN and CCC channels.

\subsubsection{Quantitative Edge and Direction Recovery}
Against the simulated-cell network, ranking each cross-cell pair by its single-gate loss reduction placed the programmed LR edge $g_{50}\to g_{60}$ first (ROC-AUC 0.93 for direct edges, 0.95 for causal paths; SI Section~S8), and edges from the sender $g_{50}$ recurred in every bootstrap resample (SI Section~S4). Within-cell edges were not identifiable from the marginals, and on this six-gene panel the greedy topologies were not significantly enriched for \emph{direct} edges. We therefore read the synthetic recovery as intercellular structure rather than the full wiring diagram.

\subsection{Single-Cell Communication Inference using Ovarian Cancer Cell Lines}\label{subsec:scrna_seq_res}
To demonstrate utility in a real biological setting, we applied QuantumXCT to an ovarian cancer single-cell transcriptomic dataset in which cancer cells interact with fibroblasts, driving their transition into cancer-associated fibroblasts (CAFs)~\cite{mori2024targeting}. In this case study, mono-cultured cells represent non-communicating baseline states, whereas co-culture represents the interacting state where CCC drives CAF polarization. We model these state shifts by learning an optimal entangling topology $U(\tau, \theta)$ that captures the joint effects of intracellular GRNs and intercellular ligand--receptor (LR) interactions.

\begin{figure}[H]
\centering
\includegraphics[width=\textwidth]{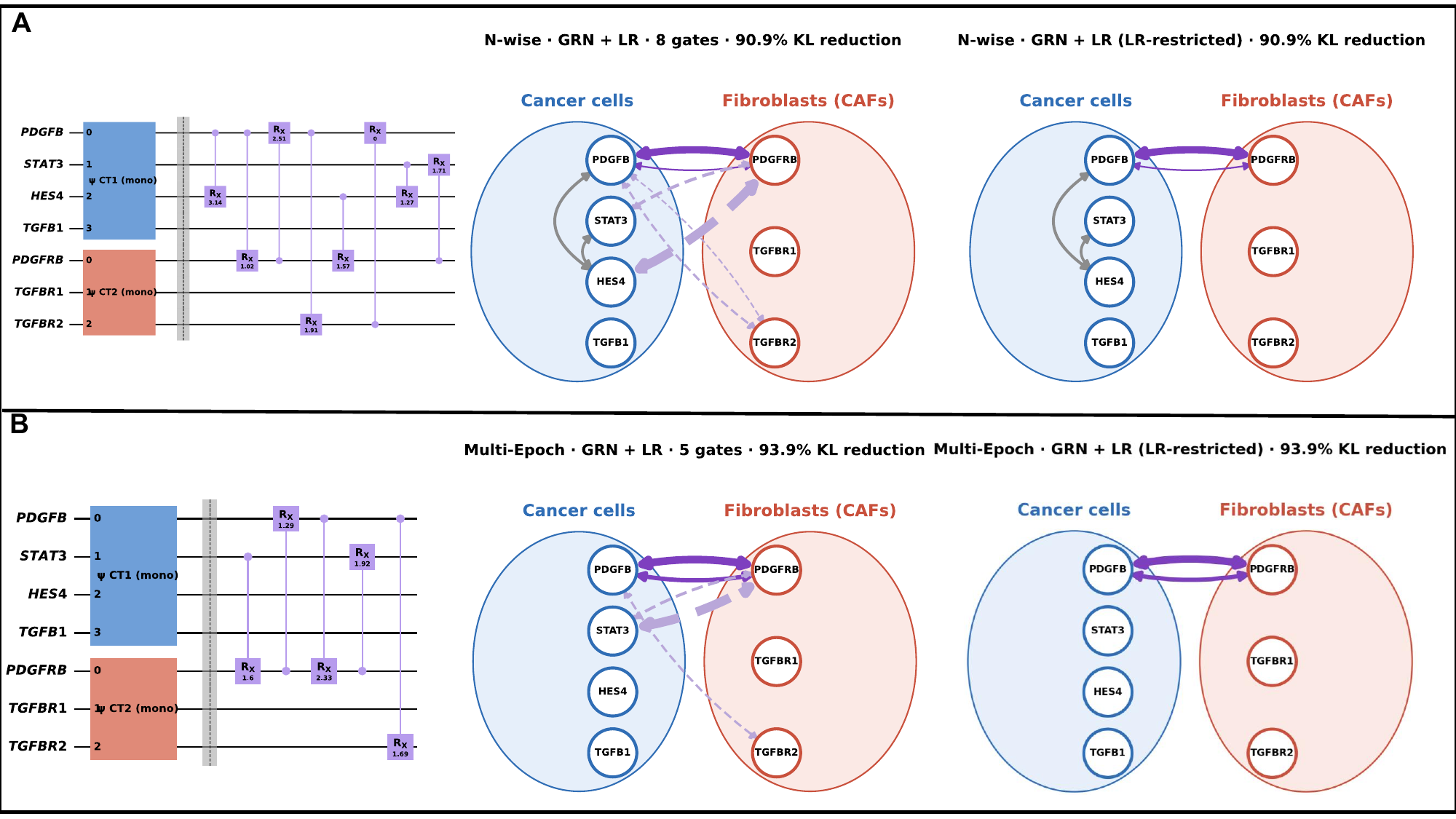}
\caption{\textbf{QuantumXCT discovers a core communication hub in an ovarian cancer dataset.} Our framework was applied to scRNA-seq data from cancer cells and fibroblasts to model their interaction.
In each row, panels display the optimized quantum circuit (left), the unrestricted network inferred by QuantumXCT (middle), and the LR-restricted network filtered by curated ligand--receptor pairs (right).
\textbf{(A) N-Wise Algorithm Results.} Convergence on an 8-gate topology identifying a core hub containing \texttt{PDGFB}, \texttt{PDGFRB}, \texttt{STAT3}, and \texttt{HES4}.
\textbf{(B) Multi-Epoch Algorithm Results.} Convergence on a 5-gate topology. Filtering candidate links against curated LR pairs demonstrates that both search heuristics recover the identical core \texttt{PDGFB-PDGFRB} intercellular communication axis.
In the networks, dark-purple solid edges denote validated ligand--receptor interactions, light-purple dashed edges indicate candidate cross-cell interactions, and grey edges represent within-cell (GRN) regulation. Edge width is proportional to the leave-one-out KL contribution, and arrows represent learned statistical relationships. All panels reflect exact statevector probabilities under $L2$ encoding.}
\label{fig:ovarian_cancer}
\end{figure}

For this purpose, we used a density-matrix cutoff ($\delta\rho>0.005$), which on this 7-gene panel retains all 42 candidate pairs, so the search is effectively exhaustive here; the cutoff reduces the search space mainly on larger panels (SI Section~S10). We then applied two primary search heuristics to identify potential CCC interactions; the results are shown in Fig.~\ref{fig:ovarian_cancer}, with a detailed breakdown of gate contributions provided in Table~\ref{tab:kl_comp_analysis}.

The \textbf{N-Wise Search} (Fig.~\ref{fig:ovarian_cancer}A) converged on an 8-gate topology that reduced the KL divergence from an initial 0.326 to 0.030 (90.9\%; exact probabilities). The resulting network correctly identified a core communication hub consisting of \texttt{PDGFB}, \texttt{PDGFRB}, \texttt{STAT3}, and \texttt{HES4}. The ablation analysis (Table~\ref{tab:kl_comp_analysis}) revealed that two interactions, \texttt{HES4 $\to$ PDGFRB} and \texttt{PDGFRB $\to$ PDGFB}, together account for essentially all of the reduction (50.7\% and 47.5\% of the total KL reduction; leave-one-out 79.1\% and 60.3\%), highlighting them as the dominant regulatory axes in this model. The biologically constrained network plot visualizes this, showing the known intercellular \texttt{PDGFB-PDGFRB} link as the primary communication channel.

The \textbf{Multi-Epoch Search} (Fig.~\ref{fig:ovarian_cancer}B) discovered a larger, 5-gate topology, and reinforced the same core biological finding. This algorithm also identified the \texttt{PDGFB-PDGFRB-STAT3} hub as the essential component, with the two interactions \texttt{STAT3$\to$PDGFRB} and \texttt{PDGFRB$\to$PDGFB} being the largest contributors (67\% of the initial KL; the circuit reduces the KL from 0.326 to 0.020 (93.9\%; exact probabilities)). Notably, this more exploratory search also proposed a cross-cell \texttt{PDGFB $\to$ TGFBR2} edge, which is not a canonical ligand--receptor pair.

Across the two searches the responding gene and the dominant link are stable: both place \texttt{PDGFRB} as the receiver and recover \texttt{PDGFRB$\to$PDGFB}, while they differ only in the control of \texttt{PDGFRB} (\texttt{HES4} for N-Wise, \texttt{STAT3} for Multi-Epoch). This is the behavior predicted by the identifiability analysis (SI Section~S3): \texttt{STAT3} and \texttt{HES4} have nearly equal basal activity in cancer cells ($\pi\approx0.68$ and $0.60$), so they act interchangeably as controls. We therefore read each edge at the level that is identified, that is, which gene responds and in which cell type, and report control identity with its bootstrap selection frequency (SI Section~S4.2).

The two-stage strategy separates structure from strength: the topology search uses fixed CRX($\pi/2$) gates, and the angles are then optimized jointly. Angles are optimized from two initializations, $\theta=0$ and $\theta=\pi/2$, to guard against local minima. Optimized from $\theta=0$ on a 5,000-shot objective, two gates of the multi-epoch topology stay at near-zero angles with KL changes within shot noise and, evaluated exactly, contribute zero at those angles; optimized from $\pi/2$, all five gates become active and the loss falls to 0.020. Table~\ref{tab:kl_comp_analysis} reports exact values for each gate: the share of the initial KL it removes, its share of the total reduction, and its isolated effect. Gate-removal analyses are given in SI Section~S4. These contributions are counterfactuals within the fitted model, not experimental interventions.

Ultimately, both the distinct algorithms converged on the same core biological insight: the cancer $\to$ CAF polarization is consistent with a CCC feedback loop involving \texttt{PDGFB}, \texttt{PDGFRB}, and \texttt{STAT3}.

\paragraph{Robustness and null models.} Restricting the circuit to within-cell gates (no inter-register gates) reduces the loss by 71\%, whereas allowing cross-cell gates reduces it by 84--88\% (Table~\ref{tab:scopes}), and the selected topology outperforms all random topologies of the same size (SI Section~S3). The hub is recovered by all three search algorithms, including the QUBO/VQE route at a stricter screen (SI Section~S10), and a cancer(gene)$\to$\texttt{PDGFRB} edge appears in every one of 50 bootstrap resamples of cells (SI Section~S4.2).

\paragraph{Independent evidence.} In the study that generated these data, \emph{PDGFRB} knockout in CAFs abolished PDGF-induced CAF activation and reduced \emph{PDGFB} expression in the cancer cells \cite{mori2024targeting}. This is the dependence encoded by the dominant edge \texttt{PDGFRB}(fibroblast)$\to$\texttt{PDGFB}(cancer). Established CCC tools also rank the PDGF and TGF-$\beta$ axes highly on the same data (SI Section~S9).

\paragraph{Post-optimization hardware validation.} Under a realistic noise model of a 127-qubit IBM device with error suppression and readout correction, the optimized circuit retained a 76.4\% loss reduction (87.8\% noiseless; SI Section~S11). On the real device (\texttt{ibm\_kingston}, 8,000 shots) the optimized circuit reduced the loss by 89.3\% ($0.162\to0.017$) with no mitigation, matching the 87.8\% noiseless reduction to within shot noise, while the no-gate baseline stayed near its encoded loss. The reduction was robust across a one-session sweep of readout correction, dynamical decoupling and twirling, and a zero-noise-extrapolation pass confirmed a noise-limited residual (SI Section~S11, Tables~S10--S11). The optimized inference layer is deliberately sparse (four cross-cell gates on seven qubits), which keeps the circuit shallow enough to recover the noiseless result on present-day hardware.

\begin{table}[htbp]

\centering
\caption{\textbf{Search scopes on ovarian and synthetic datasets (5,000-shot sampled, L1 encoding).} ``Within-cell only'' is the baseline without inter-register gates. Loss reductions are relative to baseline uncoupled loss (ovarian 0.194; synthetic 2.513). ``\#par.'' denotes the number of trainable gate parameters ($\theta$). Exact probability counterparts and null models are provided in SI Tables~S1--S2.}
\label{tab:scopes}
\small
\begin{tabular}{lrrrr}
\toprule
 & \multicolumn{2}{c}{Ovarian} & \multicolumn{2}{c}{Synthetic}\\
Model & \#par. & \% red. & \#par. & \% red.\\
\midrule
QuantumXCT, all gates (GRN + LR) & 4 & 87.8 & 13 & 75.8\\
QuantumXCT, within-cell only (GRN) & 2 & 70.6 & 6 & 21.1\\
QuantumXCT, cross-cell only (LR) & 3 & 84.4 & 9 & 76.3\\
\bottomrule
\end{tabular}
\end{table}

\begin{table}[htbp]

\centering
\caption{Gate contribution analysis for the final optimized topologies (L2 encoding, as in Fig.~\ref{fig:ovarian_cancer}). Values use exact output probabilities (no shot noise), with angles optimized from $\pi/2$. ``\% init.'': reduction at this step as a share of the initial KL. ``\% total'': the same reduction as a share of the total reduction achieved by the circuit (sums to 100\%). ``Iso.\ \%'': reduction by this gate alone, as a share of the initial KL. Gate-removal (leave-one-out) contributions and shot-noise estimates are in SI Section~S4.}
\label{tab:kl_comp_analysis}
\renewcommand{\arraystretch}{1.2}
\small
\begin{tabular}{l l l r r r r r}
\toprule
\textbf{Algorithm} & \textbf{Source} & \textbf{Target} & \textbf{$\theta_i$} & \textbf{KL} & \textbf{\% init.} & \textbf{\% total} & \textbf{Iso. \%} \\ \midrule
\multicolumn{8}{l}{\textit{\textbf{N-Wise Search}} (initial 0.326; final 0.030)} \\
& PDGFB & HES4 & 3.142 & 0.334 & $-2.3$ & $-2.5$ & $-2.3$ \\
& PDGFB & PDGFRB & 1.135 & 0.331 & 0.9 & 1.0 & 0.9 \\
& PDGFRB & PDGFB & 2.523 & 0.190 & 43.2 & 47.5 & 41.3 \\
& PDGFB & TGFBR2 & 1.909 & 0.234 & $-13.5$ & $-14.9$ & 1.2 \\
& HES4 & PDGFRB & 1.583 & 0.084 & 46.1 & 50.7 & 25.5 \\
& TGFBR2 & PDGFB & $-1.619$ & 0.084 & 0.0 & 0.0 & 0.0 \\
& STAT3 & HES4 & 1.253 & 0.079 & 1.2 & 1.4 & $-3.2$ \\
& PDGFRB & STAT3 & 1.702 & 0.030 & 15.2 & 16.8 & 9.8 \\ \midrule
\multicolumn{8}{l}{\textit{\textbf{Multi-Epoch Search}} (initial 0.326; final 0.020)} \\
& STAT3 & PDGFRB & 1.603 & 0.262 & 19.6 & 20.9 & 19.6 \\
& PDGFRB & PDGFB & 1.294 & 0.109 & 47.1 & 50.1 & 26.2 \\
& PDGFB & PDGFRB & 2.327 & 0.080 & 8.8 & 9.3 & 2.7 \\
& PDGFRB & STAT3 & 1.921 & 0.040 & 12.3 & 13.0 & 11.2 \\
& PDGFB & TGFBR2 & 1.694 & 0.020 & 6.2 & 6.6 & 1.0 \\ \bottomrule
\end{tabular}
\end{table}

\section{Discussion}\label{sec:discussion}
The emergence of single-cell transcriptomics has necessitated a shift from descriptive biology toward predictive, dynamical modeling. However, the prevailing reliance on static LR databases inherently limits the discovery of \textit{de novo} or context-specific cellular communication channels \cite{armingol2024diversification, jin2021inference}. Here, we introduce \textbf{QuantumXCT}, a hybrid quantum-classical generative framework that reframes CCC not as a search for interacting pairs, but as a learning unitary transformation between cellular probability distributions.

\subsection{Interpretability of Quantum-to-Biological Transcription}\label{subsec:disc_interpretability}
A recurring critique of QML is the ``black-box'' nature of parameterized circuits \cite{benedetti2019parameterized, cerezo2021variational}. We address this directly by demonstrating that the optimized entangling topology ($\tau_{\text{opt}}$) can be read as a structured, testable hypothesis about biological dependencies. By mapping the discovered gates back onto a biologically constrained interaction network (Fig.~\ref{fig:ovarian_cancer}), we obtain a human-interpretable ``relay map'' of intercellular interaction.

Importantly, our sequential gate contribution analysis (Table~\ref{tab:kl_comp_analysis}) allows us to ``ablate'' the model post-optimization. In the ovarian cancer model, the gates within the \texttt{PDGFB-PDGFRB-STAT3} hub account for over 93\% of the total KL reduction \cite{mori2024targeting}. Following Section~\ref{subsec:identifiability}, a gate is treated as a candidate biological interaction when it (i) exceeds null models and cell-resampling stability, (ii) possesses a distinct control activity profile, and (iii) aligns with independent evidence. \texttt{PDGFRB}$\to$\texttt{PDGFB} satisfies all three criteria, whereas the control of \texttt{PDGFRB} (\texttt{STAT3} vs.\ \texttt{HES4}, basal activities 0.68 and 0.60) remains degenerate, and within-cell edges remain exploratory. This quantitative ranking cleanly separates high-impact ``driver'' interactions from secondary ``passenger'' correlations that confound static co-expression benchmarks \cite{armingol2024diversification, dimitrov2022comparison}. Unlike classical tools such as CellChat \cite{jin2021inference, jin2025cellchat} (SI Section~S9, Table~S9) that rely on static LR co-expression probabilities, QuantumXCT quantifies the \textit{functional contribution} of each interaction to global transcriptomic state shifts, constituting the central advance of this framework (note that ``impact'' denotes state-distribution shifts rather than experimentally validated phenotypic effects).

\subsection{Algorithmic Robustness and Convergence}\label{subsec:disc_robustness}
Despite producing topologically distinct candidate circuits, the N-Wise and Multi-Epoch search heuristics (Algorithms~\ref{alg:local_search}--\ref{alg:multi_epoch}) converge on functionally equivalent representations of the core \texttt{PDGFB-PDGFRB} signaling axis \cite{hoos2004stochastic}. Multiple entangling paths can successfully encode the same high-dimensional distribution shift \cite{benedetti2019parameterized}. To distinguish stable driver interactions from near-equivalent topological artifacts, we evaluate per-edge bootstrap selection frequencies and structural invariants across cell-resampled replicates (SI Section~S4.2, Fig.~S4, Table~S5).

\subsection{Choice of Amplitude Encoding: L1 as the Default}\label{subsec:disc_encoding}
The L1 (square-root) encoding directly reproduces empirical co-culture frequencies, whereas the L2 (squared-count) encoding generates an order-2 escort distribution that emphasizes dominant cell states (Eqs.~\eqref{eq:amplitude_mo}--\eqref{eq:amplitude_l1}; Table~\ref{tab:norms}). Both encodings consistently recover the core \texttt{STAT3}$\to$\texttt{PDGFRB}$\to$\texttt{PDGFB} and \texttt{PDGFB}$\to$\texttt{TGFBR2} signaling axis (SI Section~S2, Table~S1). L2 squaring separates near-degenerate gene activities, enhancing bootstrap edge stability and adding reciprocal \texttt{PDGFB}$\leftrightarrow$\texttt{PDGFRB} links, whereas L1 preserves rare but real co-culture states. We adopt L1 as the default encoding for frequency fidelity. The synthetic circuits (Fig.~\ref{fig:sim_comparison}) are shown under L1, while the ovarian circuit figure is presented under L2 (Fig.~\ref{fig:ovarian_cancer}) for structural clarity.

\begin{table}[htbp]

\centering
\caption{L1 versus L2 amplitude encoding on the ovarian data (multi-epoch search, 5000-shot sampled, $\delta\rho>0.005$). Loss reductions are from the sampled pipeline; bootstrap frequencies are from 50 exact-scored resamples of cells.}
\label{tab:norms}
\small
\begin{tabular}{lcc}
\toprule
 & L1 (frequency) & L2 (squared counts)\\
\midrule
Loss reduction (\% of own baseline) & 87.8 & 91.5\\
\# gates & 4 & 5\\
\texttt{STAT3}$\to$\texttt{PDGFRB} (bootstrap freq.) & selected (0.34) & selected (0.80)\\
\texttt{PDGFRB}$\to$\texttt{PDGFB} & selected (0.48) & selected (1.00)\\
\texttt{PDGFB}$\to$\texttt{TGFBR2} & selected (0.54) & selected (0.82)\\
\texttt{PDGFB}$\to$\texttt{PDGFRB} & -- (0.06) & selected (0.98)\\
\texttt{PDGFRB}$\to$\texttt{STAT3} & -- (0.12) & selected (0.88)\\
\texttt{TGFBR1}$\to$\texttt{TGFB1} & selected (0.52) & -- (0.00)\\
any cancer$\to$\texttt{PDGFRB} edge & 1.00 & 1.00\\
Within-cell-only reduction & 70.6\% & 0\%\\
\bottomrule
\end{tabular}
\end{table}

\subsection{Binarization as a Biologically Motivated Encoding}
\label{subsec:disc_binarization}
Binarizing log-normalized transcriptomes into active/inactive states reflects the characteristic ON/OFF transcriptional switching observed in single-cell regulatory networks \cite{korthauer2016statistical} while accommodating transcriptomic sparsity \cite{qiu2020embracing}. Binary representations preserve cell-type-level variation and yield conclusions consistent with count-based analyses \cite{bouland2021differential, vandenBerge2023consequences}. Crucially, binarization defines the discrete \textit{topology} of the probability landscape, whereas continuous gate angles $\theta$ learn the quantitative \textit{strengths} of state transitions (Section~\ref{subsec:optimization}). While multi-level quantization (e.g., discrete expression tiers) could offer finer resolution, encoding $M$ levels per gene requires multiple qubits per gene ($k = \lceil\log_2 M\rceil$), multiplying register size and circuit depth beyond the coherence limits of current NISQ hardware \cite{preskill2018quantum}.

\subsection{Marginal Approximation, Scalability, and the Path to Full Joint
Inference}
\label{subsec:disc_marginal}
The marginal loss formulation (Eq.~\ref{eq:kl_cost}) is directly dictated by single-cell protocols, which measure cell populations independently without joint cross-cell observations \cite{kullback1951information}. By initializing the baseline as a separable tensor product $\ket{\Psi_{\text{Mo}}} = \ket{\psi^{\text{Mo}}_{\text{CT1}}} \otimes \ket{\psi^{\text{Mo}}_{\text{CT2}}}$ (Eq.~\ref{eq:initial_state_global}), QuantumXCT delegates all modeled cross-talk exclusively to the entangling unitary $U(\tau, \theta)$. Because marginal optimization alone could theoretically be satisfied without entanglement, the discovered topology represents a structured hypothesis under Section~\ref{subsec:identifiability}, supported by null-model benchmarking and cell-resampling stability (SI Section~S3). As fault-tolerant quantum hardware matures, joint state tomography can be explored to verify higher-order multi-gene dependencies that marginal objectives encode implicitly \cite{preskill2018quantum, cerezo2021variational}.

Similarly, scalability analysis of the QUBO/VQE formulation (Algorithm~3, Appendix~\ref{appendix:algo3}) highlights a NISQ-era trade-off (SI Section~S10). At strict screening thresholds ($\delta\rho>0.005$), candidate expansion (42 gate pairs) makes classical gate ordering intractable \cite{lucas2014ising}. Under a relaxed threshold ($\delta\rho>0.02$, 12 candidates), the VQE-hybrid algorithm recovers the core 2-gate communication axis (\texttt{STAT3}$\to$\texttt{PDGFRB}, \texttt{PDGFRB}$\to$\texttt{PDGFB}) with an 81.3\% KL reduction, closely tracking stochastic heuristics \cite{blumer1987occam}. While classical local searches remain the most efficient choice for NISQ devices, the QUBO architecture provides a forward-looking foundation for screening massive candidate spaces in superposition as fault-tolerant quantum devices develop \cite{farhi2014quantum, peruzzo2014variational, cerezo2021variational}.

\subsection{Model Boundaries, Classical Baselines, and Scope}\label{subsec:disc_boundaries}
QuantumXCT does not claim a computational quantum advantage over classical runtime. Instead, its core novelty lies in leveraging an \emph{entanglement-based circuit ansatz} as a quantum-inspired parameterization for modeling cross-cell communication and high-order gene dependencies \cite{romero2026qsimcells}. At matched capacity, QuantumXCT outperforms optimal transport \cite{cuturi2013sinkhorn} and tracks low-capacity baselines. 
While unconstrained classical tensor networks \cite{verstraete2004mpdo} and Born machines \cite{han2018mps} (SI Sections~S6--S7), as well as continuous flow models \cite{papamakarios2021flows, chen2018neuralode, bunne2023cellot}, fit marginal distributions tightly, they function as uninterpretable black boxes in a biological context.
Conversely, unconstrained per-gene rotations (Local-Ry, loss 0.0066) fit marginals trivially but remain biologically uninformative because they isolate per-gene re-tuning without capturing cross-talk (SI Section~S3). QuantumXCT deliberately restricts model capacity to achieve an optimal balance: using continuous gate parameters that map directly to directed, human-interpretable interaction networks \cite{huang2019qadvantage, cerezo2021variational}. While QML enhancements have been reported for 1D structural datasets \cite{xiao2023ghost} and quantum physical state learning \cite{xu2025heisenberg, hou2025measurements, garratt2024probing, kumar2026unlearnable}, fitting observational single-cell statistics provides a structured interaction hypothesis rather than definitive causal proof, a boundary shared across all observational inference tools.

This architecture operates within clear physical and methodological boundaries. Current NISQ hardware constraints \cite{preskill2018quantum} limit single-circuit deployments to pathway modules ($n_1+n_2 \leq 10$--15 qubits) due to classical state-vector simulation limits and sampling shot noise ($N_s \propto \delta^{-2}$; SI Section~S10), though larger systems can be analyzed by module tiling and scale naturally as fault-tolerant quantum hardware develops. Methodologically, QuantumXCT models a \emph{transitional} unitary mapping monoculture to co-culture states, assuming baseline basal expression rather than re-learning it. This differential framing prevents local rotations from masking true cross-talk signals, offering a natural foundation for comparative single-cell and spatial transcriptomics. Finally, while classical CCC tools (scTenifoldXct \cite{yang2023sctenifoldxct}, LIANA+ \cite{dimitrov2024liana}, CellChat \cite{jin2021inference, jin2025cellchat}; SI Section~S9) score static co-expression probabilities of known pairs, QuantumXCT provides a complementary, data-driven quantification of how strongly candidate interactions drive global transcriptomic state shifts.

\section{Conclusion}\label{sec:conclusion}
QuantumXCT provides a generative framework for inferring cell-cell communication from single-cell transcriptomics by modeling how cellular state distributions shift between non-interacting and interacting conditions. Unlike traditional approaches that rely strictly on static database matching of ligand-receptor pairs \cite{armingol2024diversification, jin2021inference}, our framework learns the quantitative, system-level impact of candidate interactions on global cell states. Whereas interpretability in QML is typically sought post hoc by explaining black-box models, QuantumXCT offers intrinsic interpretability: each entangling gate represents a named, directed, activity-gated interaction discovered directly through structure search. Our evaluations on synthetic and co-culture datasets demonstrate that sparse, capacity-restricted quantum circuits reliably capture directional signaling feedback loops, supported by null-model benchmarking, cell resampling, classical CCC tools, and published perturbation data \cite{romero2026qsimcells}. Because these inferred circuits are deliberately sparse, they remain shallow enough for NISQ-era quantum processors: physical execution on the \texttt{ibm\_kingston} device reproduced noiseless loss reductions to within shot-noise uncertainty (SI Section~S11). As quantum hardware scales beyond the NISQ era \cite{preskill2018quantum}, this framework provides a complementary, data-driven foundation for uncovering the latent interaction rules governing complex tissue microenvironments in oncology, immunology, and regenerative medicine.

\backmatter

\section*{Acknowledgments}

\section*{Declarations}
\begin{itemize}
\item\textbf{Funding}: This study was funded by the U.S. Department of Defense (DoD, GW200026) and the National Institute for Environmental Health Sciences (NIEHS, P30 ES029067) for J.J.C, Allen Endowed Chair in Nutrition \& Chronic Disease Prevention for R.S.C., and the Cancer Prevention \& Research Institute of Texas (CPRIT, RP230204) and Texas A\&M University 2026 Targeted Proposal Teams (TPT) funding program for J.J.C and R.S.C.\\
\item\textbf{Conflict of interest}: The authors have no conflict of interest.\\
\item\textbf{Data availability}: Our algorithm is publicly available on GitHub \url{https://github.com/cailab-tamu/QuantumXCT} along with the simulated data used in the study. The ovarian cancer scRNA-seq dataset used in this study is available at GSE224333. Scripts regenerating all analyses are in \texttt{python/analysis\_tools} of the repository.\\
\item\textbf{Supplementary information}: The SI contains the formal model and encoding definitions, the identifiability analysis and null models, the exact gate-contribution and bootstrap-stability analyses, the comparison with classical and tensor-network baselines, synthetic edge recovery, comparison with established CCC tools, scalability analysis, and the hardware validation.\\
\item\textbf{Author contribution}: S.R.: Conceptualization, Methodology, Formal analysis, Software, Writing - Original Draft. S.G: Investigation, Validation, Writing- Review \& Editing. R.S.C: Supervision, Resources. J.J.C: Conceptualization, Resources, Software, Supervision, Validation, Writing - Review \& Editing.
\end{itemize}

\begin{appendices}

\section{Common Algorithmic Components}
\label{appendix:common_components}
All algorithms developed for discovering the optimal entangling topology, $\tau$, share foundational components for circuit construction and evaluation. This section details these shared modules.

\subsection{Quantum Circuit Construction and Evaluation}
The evaluation of any candidate topology, $\tau$, involves constructing the corresponding quantum circuit, simulating it, and calculating a cost.

\subsubsection{Circuit Construction}\label{appendix:circuit_construction} 
A trial quantum circuit, $U(\tau)$, is assembled for a given sequence of entangling gates, $\tau = (G_k, \dots, G_1)$. The unitary operator is constructed as $U(\tau) = U_{G_k} \circ \dots \circ U_{G_1}$. During the discrete topology search phase, each gate $G_i$ in the sequence is implemented as a Controlled-RX gate with a fixed angle, CRX($\pi/2$). This strategic choice was made because a standard CNOT gate (equivalent to CRX($\pi$)) creates maximal entanglement, which might be too ``coarse" a tool for this problem. We reasoned that a discrete search using CNOTs might overlook weaker, yet biologically significant, correlations. By using CRX($\pi/2$), we allow the topology search to identify potential interactions based on their ability to create partial or ``mixed" entanglement, thereby capturing a wider range of potential inter- and intra-state interactions that minimize the cost function.

The final state of the system before measurement is $\ket{\psi'} = U(\tau)\ket{\Psi_{\text{Mo}}}$. Measurement operations are appended to each logical register (CT1 and CT2).

\subsubsection{Cost Function Evaluation}
The fitness of a circuit is quantified by its cost, $\mathcal{L}(\tau)$.
\begin{enumerate}
    \item \textbf{Quantum Simulation:} The circuit $U(\tau)$ is executed for \texttt{nshots}, yielding measurement counts $C_{\text{sim}}^{\text{CT}_k}(s)$ for each state $s$ in each logical register $k$.
    \item \textbf{Quantum Inferred Distribution Sampling:} The counts are normalized to produce marginal probability distributions from quantum sampled histograms (related to the amplitudes Eqs.~\ref{eq:amplitude_mo} and ~\ref{eq:prob_co}): $P_{\psi'}^{\text{CT}_k}(s) = C_{\text{sim}}^{\text{CT}_k}(s) / \texttt{nshots}$. The marginals are also computed exactly from the state vector; exact values are used for all contribution analyses.
    \item \textbf{KL Divergence Calculation:} The total cost is the sum of KL divergences between the simulated and target distributions (from Eq.~\ref{eq:kl_cost}):
    \begin{equation}
        D_{\text{KL}}(P_{\psi'}^{\text{CT}_k} \parallel Q_{\text{Co}}^{\text{CT}_k}) = \sum_{s} P_{\psi'}^{\text{CT}_k}(s) \log \left( \frac{P_{\psi'}^{\text{CT}_k}(s)}{Q_{\text{Co}}^{\text{CT}_k}(s)} \right)
    \end{equation}
\end{enumerate}

\section{Heuristic Pruning of the Gate Search Space}
\label{appendix:pruning}

All topology search algorithms described in this work operate not on the complete set of all possible entangling gates, but on a reduced, high-potential candidate set, $\mathcal{C}$. This initial pruning step is critical for making the subsequent combinatorial and permutational searches tractable.

We guide the search by analyzing the difference between the global density matrices of the non-interacting and interacting systems. First, we define the global state vectors for the mono-culture (initial) and co-culture (target) systems as the tensor products of their respective individual state vectors, which are derived from the experimental data as described in the main text (Eqs.~\ref{eq:amplitude_mo} and \ref{eq:prob_co}):
\begin{align}
    \ket{\Psi_{\text{Mo}}} &= \ket{\psi^{\text{Mo}}_{\text{CT1}}} \otimes \ket{\psi^{\text{Mo}}_{\text{CT2}}} \\
    \ket{\Psi_{\text{Co}}} &= \ket{\psi^{\text{Co}}_{\text{CT1}}} \otimes \ket{\psi^{\text{Co}}_{\text{CT2}}}
\end{align}
The corresponding density matrices are then defined by the outer product, $\rho = \ket{\Psi}\bra{\Psi}$. We compute the difference between the target and initial density matrices:
\begin{equation}\label{eq:dens_diff_appendix}
    \Delta\rho = \rho_{\text{Co}} - \rho_{\text{Mo}}
\end{equation}
The matrix $\Delta\rho$ highlights the specific state-to-state transitions (i.e., the off-diagonal elements) and changes in state populations (i.e., the diagonal elements) required to evolve the system from the non-interacting to the interacting state. By applying a threshold to the elements of $\Delta\rho$, we identify the most significant state-to-state transitions. From this, we generate a reduced candidate set, $\mathcal{C}$, where each element is a (control, target) qubit pair that specifies a potential CRX($\pi/2$) gate. These pairs are selected because they are the most likely to facilitate the identified transitions. This pruned set $\mathcal{C}$ serves as the direct input for all subsequent topology search algorithms.

\section{Algorithm 1: Iterative Local Search}
\label{appendix:algo1}

This appendix provides a detailed description of the \textit{Iterative Local Search} algorithm, a classical heuristic for topology discovery. The algorithm begins with the pruned candidate set $\mathcal{C}$ (Appendix~\ref{appendix:pruning}) and uses the evaluation pipeline from Appendix~\ref{appendix:common_components}. It directly addresses the combined gate selection and ordering problem by iteratively and sequentially applying three types of moves (insertion, addition, and deletion) to refine a sequence of entangling gates.

\subsection{Algorithmic Workflow: Sequential Local Search}
The algorithm operates within a main iterative loop that continues as long as any move within an iteration successfully improves the cost function. Unlike a standard local search that evaluates all neighbors before making a move, this algorithm executes three distinct phases in a fixed sequence within each iteration, immediately updating the working solution if a phase is successful. This aggressive, sequential approach is designed to rapidly descend the cost landscape, as outlined in Algorithm~\ref{alg:local_search}.

\begin{algorithm}[h!]
\caption{Iterative Local Search for Topology Discovery}
\label{alg:local_search}
\begin{algorithmic}[1]
\Statex \textbf{Input:} Candidate gate set $\mathcal{C}$, Target distributions $Q_{\text{Co}}$, Permutation length $n_{\text{choose}}$, Significance threshold \texttt{kl\_tol}.
\Statex \textbf{Output:} Optimal sequence $\tau_{\text{best}}$, Final cost $L_{\text{best}}$.
\State \textbf{Initialize:} $\tau_{\text{best}} \gets \emptyset$, $L_{\text{best}} \gets \mathcal{L}(\emptyset)$, \texttt{improvement\_made} $\gets \text{True}$.
\While{\texttt{improvement\_made}}
    \State \texttt{improvement\_made} $\gets \text{False}$
    
    \State \Comment{\textit{Phase 1: Single Gate Insertion}}
    \State $(\tau_{\text{cand}}, L_{\text{cand}}) \gets \text{FindBestSingleInsertion}(\tau_{\text{best}}, \mathcal{C}, \texttt{kl\_tol})$
    \If{$L_{\text{cand}} < (L_{\text{best}} - \texttt{kl\_tol})$}
        \State $\tau_{\text{best}} \gets \tau_{\text{cand}}$, $L_{\text{best}} \gets L_{\text{cand}}$, \texttt{improvement\_made} $\gets \text{True}$
    \EndIf
    
    \State \Comment{\textit{Phase 2: N-wise Permutation Addition (on updated sequence)}}
    \State $(\tau_{\text{cand}}, L_{\text{cand}}) \gets \text{FindBestPermutationAddition}(\tau_{\text{best}}, \mathcal{C}, n_{\text{choose}}, \texttt{kl\_tol})$
    \If{$L_{\text{cand}} < (L_{\text{best}} - \texttt{kl\_tol})$}
        \State $\tau_{\text{best}} \gets \tau_{\text{cand}}$, $L_{\text{best}} \gets L_{\text{cand}}$, \texttt{improvement\_made} $\gets \text{True}$
    \EndIf

    \State \Comment{\textit{Phase 3: Single Gate Deletion (on updated sequence)}}
    \State $(\tau_{\text{cand}}, L_{\text{cand}}) \gets \text{FindBestSingleDeletion}(\tau_{\text{best}})$
    \If{$L_{\text{cand}} < (L_{\text{best}} - \epsilon_{\text{prune}})$} \Comment{$\epsilon_{\text{prune}}$ is a small positive constant}
        \State $\tau_{\text{best}} \gets \tau_{\text{cand}}$, $L_{\text{best}} \gets L_{\text{cand}}$, \texttt{improvement\_made} $\gets \text{True}$
    \EndIf
\EndWhile

\State \textbf{Return} $\tau_{\text{best}}, L_{\text{best}}$
\end{algorithmic}
\end{algorithm}

\subsection{Phase Descriptions}
Each phase of the iteration attempts to find the best possible move of a specific type and accepts it based on a defined criterion.

\subsubsection{Phase 1: Single Gate Insertion}
This move probes the importance of gate ordering by finding the best single gate to insert and its optimal position within the current sequence, $\tau_{\text{current}}$. The search for the optimal gate-index pair $(G^*, i^*)$ is defined as:
\begin{equation}
    (G^*, i^*) = \arg\min_{G \in \mathcal{C} \setminus \tau_{\text{current}}, \, i \in [0, |\tau_{\text{current}}|]} \mathcal{L}(\text{Insert}(G, \tau_{\text{current}}, i))
\end{equation}
The resulting sequence $\tau_{\text{insert}} = \text{Insert}(G^*, \tau_{\text{current}}, i^*)$ is immediately adopted if it provides a cost reduction greater than the significance threshold, i.e., if $\mathcal{L}(\tau_{\text{insert}}) < (\mathcal{L}(\tau_{\text{current}}) - \texttt{kl\_tol})$.

\subsubsection{Phase 2: N-wise Permutation Addition}
Operating on the potentially updated sequence from the previous phase, this move considers adding a short, ordered sequence of new gates. The length of this sequence, $n$, is a user-definable parameter (\texttt{n\_choose}). This phase searches over all permutations $p$ of length $n$ drawn from the remaining candidate gates $\mathcal{C}_{\text{rem}} = \mathcal{C} \setminus \tau_{\text{current}}$. The best permutation $p^*$ to append is found by solving:
\begin{equation}
    p^* = \arg\min_{p \in \mathcal{P}_n(\mathcal{C}_{\text{rem}})} \mathcal{L}(\tau_{\text{current}} \oplus p)
\end{equation}
where $\oplus$ denotes sequence concatenation. The resulting sequence $\tau_{\text{add}} = \tau_{\text{current}} \oplus p^*$ is immediately adopted if it meets the significance threshold.

This move explicitly tests for synergistic effects between small, ordered groups of new gates. While the algorithm supports any integer $n$, the computational cost of this phase scales factorially with $n$ and polynomially with the size of $\mathcal{C}_{\text{rem}}$. Therefore, for practical applications, small values of $n$ are necessary. In this work, we set $n=2$, focusing the search on pairwise interactions, which we found provides a good balance between uncovering synergistic effects and maintaining computational tractability.

\subsubsection{Phase 3: Single Gate Deletion}
Finally, this move provides a crucial pruning mechanism to correct for earlier greedy steps. It finds the optimal index $i^*$ of a gate to remove from the current sequence by finding the move that yields the absolute lowest cost:
\begin{equation}
    i^* = \arg\min_{i \in [0, |\tau_{\text{current}}|-1]} \mathcal{L}(\text{Delete}(\tau_{\text{current}}, i))
\end{equation}
The resulting sequence, $\tau_{\text{del}} = \text{Delete}(\tau_{\text{current}}, i^*)$, is adopted if its cost is lower than the current cost by more than a small, fixed margin ($\mathcal{L}(\tau_{\text{del}}) < \mathcal{L}(\tau_{\text{current}}) - \epsilon_{\text{prune}}$). This less strict criterion allows the algorithm to remove redundant gates to favor parsimony, even for marginal cost improvements.

\section{Algorithm 2:  Multi-Epoch Sequential Construction}
\label{appendix:algo2}

This appendix provides a detailed description of the \textit{Multi-Epoch Sequential Construction} algorithm. This heuristic is designed as a faster, stochastic alternative to the exhaustive local search, aiming to balance a broad exploration of the solution space with a final selection principle that favors model parsimony. The algorithm begins with the pruned candidate set $\mathcal{C}$ (Appendix~\ref{appendix:pruning}) and utilizes the evaluation pipeline from Appendix~\ref{appendix:common_components}.

\subsection{Algorithmic Workflow}
The algorithm is composed of three interconnected phases. First, it explores the solution space via multiple \textbf{Forward Greedy Construction} epochs; an epoch only proceeds if the initial random gate improves upon the baseline. Second, an \textbf{Immediate Backward Greedy Refinement} is triggered \textit{if} a forward search yields a new best-so-far solution; this phase removes irrelevant gates to ensure each addition in the sequence is strictly necessary. Finally, after all epochs are complete, a global selection phase based on Occam's Razor determines the final result. The process is outlined in Algorithm~\ref{alg:multi_epoch}.

\begin{algorithm}[h!]
\caption{Multi-Epoch Construction with Path Refinement and Occam's Razor}
\label{alg:multi_epoch}
\begin{algorithmic}[1]
\Statex \textbf{Input:} Candidate gate set $\mathcal{C}$, Target distributions $Q$, Epochs $N_{\text{epochs}}$, Threshold $\epsilon$ (\texttt{kl\_tol}).
\Statex \textbf{Output:} Optimal gate sequence $\tau^*$, Final cost $L^*$.
\State \textbf{Initialize:} $\tau_{\text{best}} \gets \emptyset$, $L_{\text{best}} \gets \mathcal{L}(\emptyset)$, $\mathcal{H}_{\text{eval}} \gets \{(L_{\text{best}}, \emptyset)\}$, $L_{\emptyset} \gets L_{\text{best}}$.
\State $\mathcal{C}_{\text{shuffled}} \gets \text{Shuffle}(\mathcal{C})$

\State \Comment{\textit{Phase 1: Multi-Epoch Exploration \& Refinement}}
\For{epoch from $1$ to $\min(N_{\text{epochs}}, |\mathcal{C}|)$}
    \State $G_{\text{start}} \gets \mathcal{C}_{\text{shuffled}}[\text{epoch}]$
    \State $L_{\text{start}} \gets \mathcal{L}((G_{\text{start}}))$
    
    \If{$L_{\text{start}} < L_{\emptyset}$} \Comment{Proceed only if $G_{\text{start}}$ improves on baseline}
        \State \Comment{Step 1.1: Greedy Forward Construction}
        \State $(\mathcal{H}_{\text{path}}, \tau_{\text{path}}, L_{\text{path}}) \gets \text{GreedyForward}((G_{\text{start}}), \mathcal{C})$
        \State $\mathcal{H}_{\text{eval}} \gets \mathcal{H}_{\text{eval}} \cup \mathcal{H}_{\text{path}}$
        
        \If{$L_{\text{path}} < L_{\text{best}}$} \Comment{New global champion found on this path}
            \State $L_{\text{best}} \gets L_{\text{path}}, \tau_{\text{best}} \gets \tau_{\text{path}}$
            
            \State \Comment{Step 1.2: Immediate Greedy Removal (Refinement)}
            \State $(\mathcal{H}_{\text{rem}}, \tau_{\text{rem}}, L_{\text{rem}}) \gets \text{GreedyRemoval}(\tau_{\text{best}}, L_{\text{best}}, \epsilon)$
            \State $\mathcal{H}_{\text{eval}} \gets \mathcal{H}_{\text{eval}} \cup \mathcal{H}_{\text{rem}}$
            
            \If{$L_{\text{rem}} < L_{\text{best}}$}
                \State $L_{\text{best}} \gets L_{\text{rem}}, \tau_{\text{best}} \gets \tau_{\text{rem}}$
            \EndIf
        \EndIf
    \EndIf
\EndFor

\State \Comment{\textit{Phase 2: Global Selection via Occam's Razor}}
\State Sort $\mathcal{H}_{\text{eval}}$ by sequence length $|\tau|$ in ascending order.
\State $(\tau^*, L^*) \gets \text{First element of sorted } \mathcal{H}_{\text{eval}}$
\For{each $(L_i, \tau_i) \in \text{sorted } \mathcal{H}_{\text{eval}}$}
    \If{$L_i < (L^* - \epsilon)$} \Comment{Accept longer sequence only if gain $> \epsilon$}
        \State $\tau^* \gets \tau_i, L^* \gets L_i$
    \EndIf
\EndFor

\State \textbf{Return:} $\tau^*, L^*$
\end{algorithmic}
\end{algorithm}

\subsection{Phase Descriptions}

\subsubsection{Phase 1: Multi-Epoch Forward Construction and Refinement}
This phase explores distinct regions of the solution space by initiating multiple independent, stochastic greedy searches. For each epoch, a candidate gate $G_{start}$ is selected from the shuffled set $\mathcal{C}_{shuffled}$. To maintain computational efficiency, a full forward greedy search is only initiated if this initial single-gate sequence improves upon the baseline empty circuit: 
\begin{equation}
\mathcal{L}((G_{start})) < \mathcal{L}(\emptyset)
\end{equation}

\begin{itemize}
    \item \textbf{Forward Greedy Construction:} If the starting gate is promising, the algorithm performs an iterative constructive search. At each step $k$, it identifies the optimal gate to append from the candidate set:
    \begin{equation}
    G^*_{k+1} = \arg\min_{G \in \mathcal{C} \setminus \tau_k} \mathcal{L}(\tau_k \oplus G)
    \end{equation}
    The new sequence $\tau_{k+1} = \tau_k \oplus G^*_{k+1}$ is accepted if $\mathcal{L}(\tau_{k+1}) < \mathcal{L}(\tau_k)$. This process continues until no single gate addition can improve the cost or the maximum depth is reached. The entire history of evaluated sequences is recorded in $\mathcal{H}_{eval}$.

    \item \textbf{Immediate Backward Refinement:} If the forward search yields a sequence $\tau_{path}$ that is a new global champion ($\mathcal{L}(\tau_{path}) < L_{best}$), a refinement step is triggered to prune redundant gates. Starting with $\tau_0 = \tau_{path}$, the algorithm evaluates the removal of every gate in the current sequence to identify the position $i^*$ that yields the minimum cost:
    \begin{equation}
    i^* = \arg\min_{i \in [0, |\tau_k|-1]} \mathcal{L}(\text{Delete}(\tau_k, i))
    \end{equation}
    The pruned sequence $\tau_{k+1} = \text{Delete}(\tau_k, i^*)$ is accepted if and only if $\mathcal{L}(\tau_{k+1}) < \mathcal{L}(\tau_k) - \delta$, where $\delta = 0.3 \cdot \epsilon$ represents a significance margin. This greedy positional removal continues iteratively until no single gate deletion satisfies the improvement criterion or the minimum depth is reached. This refinement phase ensures that any redundant gates introduced during the constructive phase are eliminated.
\end{itemize}

\subsubsection{Phase 2: Final Selection (Occam's Razor)}
To select the final topology $\tau^*$, the algorithm applies a model parsimony principle. The complete history of all unique sequences $\mathcal{H}_{eval}$ evaluated during all forward and backward phases is collected and sorted by sequence length $|\tau|$ in ascending order. 

The algorithm iterates through this sorted list, starting with the simplest model (the empty circuit). A more complex (longer) sequence $\tau_i$ is only accepted as the new best solution if its cost provides an improvement greater than the significance threshold $\epsilon$ (\texttt{kl\_tol}):
\begin{equation}
\mathcal{L}(\tau_i) < \mathcal{L}(\tau^*) - \epsilon
\end{equation}
This ensures that the final selected topology $\tau^*$ is the most parsimonious model that achieves a significant reduction in cost, effectively preventing over-parameterization and ensuring model robustness.

\section{Algorithm 3: QUBO-based Variational Selection}
\label{appendix:algo3}
This appendix details our quantum-native approach, which begins with the pruned candidate set $\mathcal{C}$ (Appendix~\ref{appendix:pruning}). This strategy reframes the gate selection problem as a Quadratic Unconstrained Binary Optimization (QUBO) problem, decoupling it into a quantum-assisted gate selection and a final classical ordering stage.

\subsection{Stage 1: KL-QUBO Formulation for Gate Selection}
The goal of this stage is to map the circuit optimization problem onto a QUBO framework. We define a binary vector $\mathbf{x} \in \{0, 1\}^N$, where $x_i = 1$ if the candidate CRX gate $G_i \in \mathcal{C}$ is included in the circuit for a qubit pair, and $x_i = 0$ otherwise. The objective is to minimize the energy function:
\begin{equation}
    E(\mathbf{x}) = \sum_{i=1}^N Q_{ii} x_i + \sum_{i < j} Q_{ij} x_i x_j
\end{equation}

The construction of the QUBO matrix $Q$ proceeds in two steps:

\subsubsection{KL Divergence Matrix Construction}
We pre-compute a KL divergence matrix $M_{KL} \in \mathbb{R}^{N \times N}$ by evaluating the cost of single gates and gate pairs against the target distributions. Let $\mathcal{L}(\tau)$ denote the summed KL divergence of a circuit containing the sequence of gates $\tau$. The elements of $M_{KL}$ are defined as:
\begin{equation}
    (M_{KL})_{ij} = 
    \begin{cases} 
      \mathcal{L}(G_i) & \text{if } i = j \\
      \mathcal{L}(G_i, G_j) & \text{if } i \neq j
    \end{cases}
\end{equation}
where $L_{\emptyset} = \mathcal{L}(\emptyset)$ represents the baseline KL divergence of the circuit with no additional CRX gates.

\subsubsection{QUBO Matrix Derivation}
The matrix $M_{KL}$ is transformed into the QUBO matrix $Q$ by isolating the individual and synergistic contributions of the gates:

\begin{enumerate}
    \item \textbf{Linear Coefficients (Diagonal Terms):} The diagonal elements $Q_{ii}$ represent the improvement (or degradation) a single gate provides relative to the baseline:
    \begin{equation}
        Q_{ii} = (M_{KL})_{ii} - L_{\emptyset}
    \end{equation}
    
    \item \textbf{Quadratic Coefficients (Off-Diagonal Terms):} For any pair of gates $\{G_i, G_j\}$, the physical circuit must adopt a specific temporal order. To simplify the optimization, we assume the solver utilizes the optimal ordering. We first define the best-order cost:
    \begin{equation}
        L_{best}(i,j) = \min \left( (M_{KL})_{ij}, (M_{KL})_{ji} \right)
    \end{equation}
    The interaction term $Q_{ij}$ is then derived by subtracting the individual costs from the joint cost to isolate the synergy:
    \begin{equation}
        Q_{ij} = (L_{best}(i,j) - L_{\emptyset}) - Q_{ii} - Q_{jj}
    \end{equation}
\end{enumerate}

Any sequence that results in an infinite or non-computable KL divergence is assigned a penalty $P \gg |Q_{ij}|_{max}$ to ensure the solver avoids invalid gate configurations.

\subsection{Stage 2: Variational Quantum Solution}
The QUBO is converted to an Ising Hamiltonian, whose ground state (the optimal gate set) is found using VQE or QAOA.
\begin{itemize}
    \item \textbf{VQE:} We use an \texttt{EfficientSU2} ansatz with a classical COBYLA optimizer to find the ground state.
    \item \textbf{QAOA:} The same QUBO formulation can be addressed with QAOA; we did not run QAOA in this work and note it only as a natural alternative.
\end{itemize}
The output of the quantum solver is a probability distribution over possible solutions. We take the top-k most probable bitstrings as candidate solutions.

\subsection{Stage 3: Hybrid Post-Processing}
The quantum solver provides an optimal \textit{set} of gates, but not their order. For each of the top-k gate sets identified, we perform a final classical search to find the optimal permutation. This search is conducted using our \textbf{Multi-Epoch Sequential Construction (Algorithm 2)}, which efficiently finds a high-quality ordering for the small, promising set of gates selected by the quantum algorithm. The best sequence found across all top-k VQE solutions is chosen as the final topology.

\section{Spatial Mechanistic Simulation Framework}
\label{appendix:spatial_sim_cells}

To benchmark the quantum information-theoretic algorithm's ability to resolve directional causality, we developed a synthetic tissue framework that explicitly separates spatial signaling from intracellular regulation.

\subsection{Spatial Architecture and Interaction Kernel}
The tissue comprises $N$ cells $\mathcal{C}$ with spatial coordinates $P_i \in \mathbb{R}^2$. Intercellular connectivity is governed by a Gaussian decay kernel:
\begin{equation}
W_{ij} = \exp\left( -\frac{\|P_i - P_j\|^2}{2\sigma^2} \right), \quad i \neq j
\end{equation}
Cells are partitioned into four scenarios: \textit{Interacting Pairs} (A and B clusters within signaling range), \textit{Lone Senders} ($A_{lone}$), and \textit{Lone Receivers} ($B_{lone}$). This setup allows for the evaluation of algorithm robustness against spatial co-occurrence false positives.

\subsection{The Latent Manifold and Causality Chain}
The simulation models a continuous latent state $\mathbf{x}_i \in \mathbb{R}^G$ representing log-transformed transcriptomic potential. Every gene is initialized at a stochastic baseline $\bar{x}_{g,i} \sim \mathcal{N}(\mu_0, \sigma_{noise})$, where $\mu_0 = -2.0$ represents ``Biological Zero.'' Causality advances through a rigid temporal sequence:

\begin{enumerate}
    \item \textbf{Ligand Activation:} Sender cell ligands $x_{l,j}$ are manually shifted upward from baseline to represent active signal secretion.
    \item \textbf{Intercellular Signaling ($S_{r,i}$):} Receiver cell $i$ integrates neighbor signals by bridging the source expression ($x_{l,j}$) and induced response ($x_{r,i}$):
    \begin{equation}
    x_{r,i} = \bar{x}_{r,i} + \log\left( 1 + \alpha \sum_{j \in \mathcal{C}_{sender}} W_{ij} \cdot \exp(x_{l,j}) \right)
    \end{equation}
    \item \textbf{Intracellular GRN Propagation:} Downstream targets respond to the receptor via a GRN adjacency matrix $A_{gk}$:
    \begin{equation}
    x_{g,i} = x_{g,i} + \sum_{k \in \mathcal{G}} A_{gk} \cdot (x_{k,i} - \mu_0) \cdot \mathbb{I}(x_{k,i} > \tau)
    \end{equation}
    By subtracting $\mu_0$, we ensure the network responds only to relative activation (the ``shout'') while ignoring baseline stochasticity (the ``hum'').
\end{enumerate}

\subsection{Causal Benchmarking Scenarios}\label{appendix:lr_grn_benchmark}
We implement two distinct circuits to challenge the algorithm's resolution:
\begin{itemize}
    \item \textbf{Feedback Loop:} A three-stage intercellular circuit:
    \begin{equation}\label{appendix:lr_activation_path}
    \underbrace{50_{L} \xrightarrow{W_{ij}} 60_{R}}_{\text{Forward Signaling}} \implies \underbrace{60 \to 70 \to \{71, 72, 80_{L}\}}_{\text{Reactive GRN}} \implies \underbrace{80_{L} \xrightarrow{W_{ji}} 90_{R}}_{\text{Retrograde Feedback}}.
    \end{equation}
    \item \textbf{Autonomous Module:} An independent module ($73 \to \{74, 75\}$) initialized globally to test the decoupling of spatial information from intrinsic mutual information.
\end{itemize}

\subsection{Measurement Model and Sparsity}
To simulate technical capture inefficiency, the latent manifold is projected into observed UMI counts $Y_{gi}$ via a Negative Binomial distribution with Bernoulli-mask dropout:
\begin{equation}
Y_{gi} \sim \text{NB}\left(\exp(x_{gi}), \phi\right) \cdot \text{Bernoulli}(1 - \pi)
\end{equation}
With $\pi = 0.4$, the environment features $\sim 90\%$ sparsity, forcing the algorithm to distinguish biological signal from technical noise.

\noindent\textbf{Implementation Note:} The complete Python source code is available at \href{https://github.com/cailab-tamu/QuantumXCT/tree/main/python/simCellsSpace}{github.com/cailab-tamu/QuantumXCT/python/simCellsSpace}.

\end{appendices}

\bibliography{sn-bibliography}

\clearpage
\setcounter{section}{0}\setcounter{figure}{0}\setcounter{table}{0}
\setcounter{equation}{0}\setcounter{definition}{0}\setcounter{lemma}{0}
\renewcommand{\thesection}{S\arabic{section}}
\renewcommand{\thefigure}{S\arabic{figure}}
\renewcommand{\thetable}{S\arabic{table}}
\renewcommand{\theequation}{S\arabic{equation}}
\renewcommand{\thedefinition}{S\arabic{definition}}
\renewcommand{\thelemma}{S\arabic{lemma}}
\makeatletter
\renewcommand{\theHsection}{S\arabic{section}}
\renewcommand{\theHfigure}{S\arabic{figure}}
\renewcommand{\theHtable}{S\arabic{table}}
\renewcommand{\theHequation}{S\arabic{equation}}
\makeatother

\begin{center}
{\Large Supplementary Information for ``QuantumXCT: Learning Interaction-Induced State Transformation in Cell-Cell Communication via Quantum Entanglement and Generative Modeling''\par}
\bigskip
{\large Selim Romero, Shreyan Gupta, Robert S. Chapkin, James J. Cai\par}
\end{center}
\bigskip


\section{Notation and the QuantumXCT Model}\label{si:model}
Let cell type 1 (CT1) be described by $n_1$ selected genes and cell type 2 (CT2) by $n_2$ genes, with $N=n_1+n_2$ qubits in total. After binarization, each cell of CT$k$ is a bit-string $s\in\{0,1\}^{n_k}$. We write $C^{c}_{k}(s)$ for the number of cells of CT$k$ in condition $c\in\{\Mo,\Co\}$ with pattern $s$ (the non-interacting condition $\Mo$, a.k.a.\ mono-culture, and the interacting condition $\Co$, a.k.a.\ co-culture), and $\hat p^{c}_k(s)=C^c_k(s)/\sum_{s'}C^c_k(s')$ for the empirical frequency. Qubit $j$ of register $k$ encodes gene $j$ of CT$k$; register 1 occupies qubits $0,\dots,n_1-1$ and register 2 qubits $n_1,\dots,N-1$. Bit-strings are written most-significant-bit first, with qubit~0 as the rightmost character (the Qiskit convention); the global basis index is $x = x_2\,2^{n_1}+x_1$, where $x_1$ and $x_2$ are the integer basis indices of registers 1 and 2.

Given an encoding map $\alpha(\cdot)$ (Section~\ref{si:encoding}), the model is
\begin{equation}
\ket{\psi(\tau,\theta)} = U(\tau,\theta)\,\bigl(\ket{\alpha(C^{\Mo}_1)}\otimes\ket{\alpha(C^{\Mo}_2)}\bigr),\qquad
U(\tau,\theta)=\prod_{\ell=L}^{1}\mathrm{CRX}_{c_\ell\to t_\ell}(\theta_\ell),
\label{eq:si_model}
\end{equation}
where $\tau=((c_1,t_1),\dots,(c_L,t_L))$ is an ordered list of (control, target) qubit pairs and
$\mathrm{CRX}_{c\to t}(\theta)=\ket0\!\bra0_c\otimes I_t+\ket1\!\bra1_c\otimes e^{-i\theta X_t/2}$.
Let $P_k(\tau,\theta)$ be the computational-basis marginal of register $k$ and $Q_k$ the target (co-culture) distribution in the same encoding. The loss is
\begin{equation}
\mathcal L(\tau,\theta)=\sum_{k=1}^{2}\KL\bigl(P_k(\tau,\theta)\,\|\,Q_k\bigr),\qquad
\KL(P\|Q)=\sum_{s}\tilde P(s)\log\frac{\tilde P(s)}{\tilde Q(s)},
\label{eq:si_loss}
\end{equation}
with $\tilde P=\max(P,\varepsilon)$, $\tilde Q=\max(Q,\varepsilon)$ and $\varepsilon=10^{-9}$, exactly as implemented. Note that the model distribution is the \emph{first} argument (a mode-seeking, ``reverse'' KL): the model is penalized for placing probability on patterns that are absent in co-culture.

\paragraph{Exact and sampled estimators.} $P_k$ is evaluated either exactly from the state vector or from $N_s=5,000$ shots of a simulator (the sampled estimator $\hat P_k$). The contribution analyses and tables in this SI use the exact values; the sampled estimator is used to quantify shot noise (Section~\ref{si:shot}). The main-text figure panels and network edge widths instead use the sampled estimator, as stated in their captions.

\section{Amplitude Encodings: L1 versus L2}\label{si:encoding}
\begin{definition}[Encodings]
For counts $C(s)\ge0$,
\begin{align}
\text{L2 (squared-count):}\quad & \alpha^{(2)}_s=\frac{C(s)}{\|C\|_2},\qquad |\alpha^{(2)}_s|^2=\frac{C(s)^2}{\sum_{s'}C(s')^2},\label{eq:si_l2}\\
\text{L1 (square-root, default):}\quad & \alpha^{(1)}_s=\sqrt{\frac{C(s)}{\|C\|_1}},\qquad |\alpha^{(1)}_s|^2=\hat p(s).\label{eq:si_l1}
\end{align}
\end{definition}
The L1 encoding is the unique non-negative amplitude vector whose Born distribution equals the empirical frequency. The L2 Born distribution is the order-2 escort distribution $\hat p^{(2)}(s)=\hat p(s)^2/\sum_{s'}\hat p(s')^2$: it sharpens the empirical distribution, up-weighting common patterns and down-weighting rare ones (a pattern with half the frequency of another receives a quarter of its probability). Because the same map is applied to the initial state, to the target and to the density-matrix screening, the loss~\eqref{eq:si_loss} compares like with like in either encoding. The L2 loss, however, is a loss on the escort distributions, not on the observed frequencies. The two encodings share the same support, so the patterns that are present or absent are identical; they differ only in how probability mass is weighted.

Table~\ref{tab:si_norms} compares the two encodings on the ovarian data with the same search (multi-epoch, exact scoring), run in three candidate scopes, \emph{intra} (within-cell, GRN-only), \emph{inter} (cross-cell, ligand-receptor) and \emph{all} (combined GRN+LR), defined in Section~\ref{si:scopes}. The core axis \texttt{STAT3}$\to$\texttt{PDGFRB}, \texttt{PDGFRB}$\to$\texttt{PDGFB} consists of cross-cell edges and is captured in the combined (all) scope under both encodings, not in the within-cell (intra) scope. Both encodings also place a cross-cell \texttt{PDGFB}$\to$\texttt{TGFBR2} edge, which is not a canonical ligand--receptor pair. L2 adds the reciprocal \texttt{PDGFB}$\to$\texttt{PDGFRB} and \texttt{PDGFRB}$\to$\texttt{STAT3} links, whereas L1 adds \texttt{TGFBR1}$\to$\texttt{TGFB1}. With L2, no within-cell (GRN-only) topology improves the loss, whereas L1 reduces it by 71\%: the escort distribution concentrates on the dominant patterns, which within-cell rotations cannot redistribute. Because L1 models the measured frequencies directly and keeps rare but real co-culture states, we adopt it as the default and discuss the choice in the main text.

\begin{table}[htbp]
\centering\small
\caption{\textbf{L1 versus L2 encoding, ovarian cancer--fibroblast data, exact scoring, multi-epoch search ($\delta\rho>0.005$, selection tolerance $0.01$).} KL values are in each encoding's own space; ``shared'' lists edges common to the L1 and L2 all-scope topologies.}\label{tab:si_norms}
\setlength{\tabcolsep}{3pt}\begin{tabular}{llrrrp{4.9cm}}
\toprule
Encoding & Scope & \#gates & KL$_0$ & KL$_{\rm final}$ (\% red.) & Topology \\
\midrule
L1 & all   & 4 & 0.194 & 0.021 (89.0) & STAT3$\to$PDGFRB, PDGFRB$\to$PDGFB, PDGFB$\to$TGFBR2, TGFBR1$\to$TGFB1\\
L1 & intra & 2 & 0.194 & 0.056 (71.1) & STAT3$\to$PDGFB, PDGFB$\to$STAT3\\
L1 & inter & 5 & 0.194 & 0.015 (92.1) & HES4$\to$PDGFRB, PDGFRB$\to$PDGFB, TGFBR2$\to$STAT3, TGFB1$\to$TGFBR2, TGFBR1$\to$TGFB1\\
L2 & all   & 5 & 0.326 & 0.020 (93.9) & STAT3$\to$PDGFRB, PDGFRB$\to$PDGFB, PDGFB$\to$PDGFRB, PDGFRB$\to$STAT3, PDGFB$\to$TGFBR2\\
L2 & intra & 0 & 0.326 & 0.326 (0.0) & --\\
L2 & inter & 5 & 0.326 & 0.020 (93.9) & identical to L2 all\\
\midrule
\multicolumn{6}{p{12cm}}{Shared L1$\cap$L2 (all): STAT3$\to$PDGFRB, PDGFRB$\to$PDGFB, PDGFB$\to$TGFBR2 (Jaccard 0.50).}\\
\bottomrule
\end{tabular}
\end{table}

\section{Identifiability from Marginal Distributions}\label{si:ident}
Co-culture and mono-culture experiments measure the two cell populations separately, so no joint distribution over (CT1, CT2) patterns exists; the loss~\eqref{eq:si_loss} constrains only the two marginals. Matching both marginals does not by itself require communication: arbitrary single-register unitaries can already reproduce any pair of target marginals. What makes a QuantumXCT edge interpretable is the restricted model class, not the quality of the fit.

This section develops what a QuantumXCT edge can and cannot identify in general terms; the specific topologies it refers to are those reported in the main-text Results. The SI is intended to be read alongside the main text.

The design encodes a single hypothesis. The mono-culture state already carries each gene's basal activity and the within-cell regulatory structure. The model adds \emph{no} free single-qubit rotations: every change between mono- and co-culture must be produced by a controlled rotation whose control is a panel gene, and is thereby attributed either to a partner-cell gene (a cross-register edge, a candidate cell--cell interaction) or to a same-cell gene (a within-register edge, a candidate GRN rewiring). An edge is kept only when a gene differs between mono- and co-culture. If a gene behaves the same in both conditions the difference is negligible from the start, so an edge adds nothing useful; through the CRX gate it can only double or flip that gene's signal and raise the loss, so the pruning removes it. The informative signal is instead a shift the baseline cannot produce on its own, captured by an edge, within-cell (GRN) or cross-cell (LR), and kept only if it lowers the loss. A gene whose activity is similar between the two conditions therefore contributes no edge unless it changes strongly.

A single cross-register gate changes only the marginal of its \emph{target} register. Its effect on the target is scaled by how often the control gene is on, its basal activity $\pi_c$ (the fraction of cells in which it is on), because the CRX rotation acts only on the control's active branch. This fixes what an edge identifies: (i) the responding cell type is identifiable, since only the target side changes; (ii) the control gene is identifiable only up to its activity profile: two candidate controls with nearly the same activity $\pi_c$ act almost identically on the target, so they cannot be told apart from the data alone, and a control that is almost always on acts as a near-unconditional rotation; (iii) in multi-gate circuits later gates depend on earlier ones, adding order-dependent information but not removing (i)--(ii). \emph{Direction versus entanglement.} The inferred direction of an edge is operational. A CRX gate alters only the target register's computational-basis marginal and leaves the control's marginal unchanged, so the direction marks which side responds. The state, its marginals and the resulting distribution are not symmetric under swapping control and target; only the entanglement the gate creates is symmetric, shared equally by the two qubits, and carries no direction of its own. Reciprocal edges are therefore read as mutual coupling rather than one-way influence, consistent with the directed-but-symmetric structure formalized by quantum causal models \cite{allen2017quantum}.

We interpret QuantumXCT edges under three explicit assumptions.
\begin{itemize}
\item[(A1)] \emph{Attribution.} With no free single-qubit rotations, every mono$\to$co change is carried by a controlled rotation on a panel gene: a cross-register edge (candidate CCC) or a within-register edge (candidate GRN rewiring).
\item[(A2)] \emph{Parsimony.} Among topologies of similar loss, the smallest is preferred. The multi-epoch search enforces this globally through its Occam's-razor selection; the N-Wise search approximates it by stopping once added gates no longer substantially lower the loss and pruning redundant ones.
\item[(A3)] \emph{Activity-gated response.} An edge $c\to t$ reads as ``the co-culture shift of gene $t$ is best explained as a response gated by the activity of gene $c$''; it is not a claim of molecular binding or causation. Controls with indistinguishable activity are resolved with prior knowledge or orthogonal experiments.
\end{itemize}
Under (A1)--(A3), a QuantumXCT edge is a data-driven hypothesis about which gene's activity in one population best explains a distributional change in the other; it becomes biological evidence only when supported by independent data (Section~\ref{si:ccc}).

\begin{figure}[h]
\centering
\includegraphics[width=\textwidth]{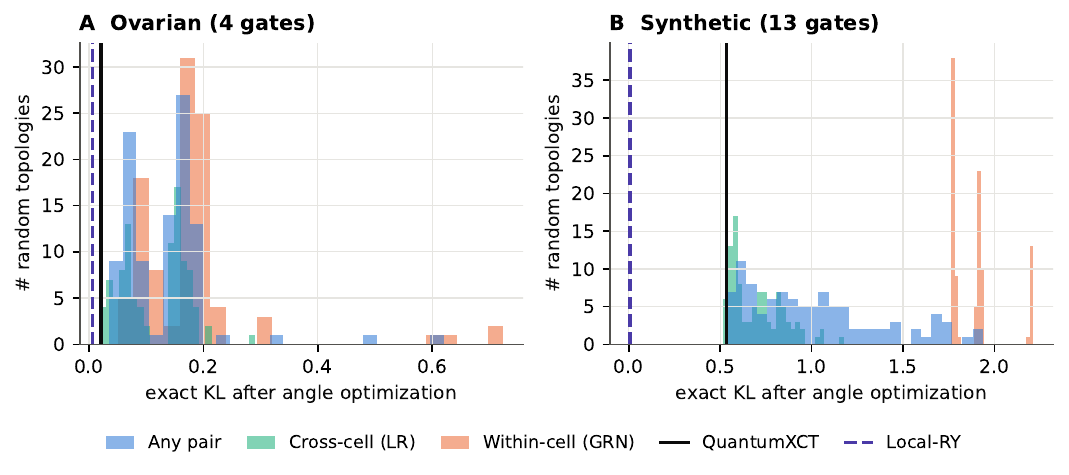}
\caption{\textbf{QuantumXCT versus same-size random topologies.} Exact loss, after angle optimization, of 100 random topologies with the same number of gates as QuantumXCT, drawn from all gene pairs (blue), cross-cell pairs only (green) or within-cell pairs only (orange). The solid black line marks the QuantumXCT topology and the dashed violet line the local-RY null (free single-qubit rotations only, no two-qubit gates; 7 parameters for ovarian, 6 for synthetic). (A) ovarian, 4 gates; (B) synthetic, 13 gates.}\label{fig:si_null}
\end{figure}

\paragraph{Null models.} We test the inferred topologies against two kinds of null (Table~\ref{tab:si_nulls}). The first is a within-cell-only topology with no cross-cell gates (the \emph{intra} scope; Table~\ref{tab:si_norms}, rows ``intra''), which asks whether cross-cell edges are needed at all. The second is a random null (Table~\ref{tab:si_random}, Fig.~\ref{fig:si_null}): for each dataset we draw 100 topologies with the same number of gates as the selected QuantumXCT topology ($L=4$ ovarian, $L=13$ synthetic) and fit each one's angles with the same optimizer and objective used for QuantumXCT, so every random topology is given its best possible fit and only the choice of edges differs. The $L$ gates are drawn three ways: from all gene pairs, from cross-cell pairs only, or from within-cell pairs only. A free per-gene rotation model with no interactions (``local self-regulation'') is reported as a reference point in the method comparison of Section~\ref{si:budget}, since it models re-tuning rather than communication.

\begin{table}[htbp]
\centering\small
\caption{\textbf{Exact loss of the QuantumXCT topology (saved L1 run, 5,000-shot optimized angles) versus reference models.} All values are exact KL (lower is better); $L$ is the number of interaction gates in the selected topology.}\label{tab:si_nulls}
\begin{tabular}{p{7cm}rr}
\toprule
 & Ovarian ($L=4$) & Synthetic ($L=13$)\\
\midrule
Baseline (no gates) & 0.194 & 2.513\\
QuantumXCT, best-effort topology (exact loss) & 0.0212 & 0.533\\
Local-RY, one free rotation per qubit (no two-qubit gates) & 0.0066 & 0.0071\\
\bottomrule
\end{tabular}
\end{table}

\begin{table}[htbp]
\centering\small
\caption{\textbf{Random-topology null.} For each dataset, the median and minimum exact loss over 100 random topologies of the same size as the selected QuantumXCT topology ($L=4$ ovarian, $L=13$ synthetic), each fit with the same optimizer and objective. ``Frac.\ $\le$'' is the fraction that match or beat QuantumXCT (0.0212 ovarian, 0.533 synthetic); a small value means the specific edges, not merely the gate count, drive the fit (interpreted below).}\label{tab:si_random}
\setlength{\tabcolsep}{5pt}\begin{tabular}{l rrr rrr}
\toprule
 & \multicolumn{3}{c}{Ovarian ($L=4$)} & \multicolumn{3}{c}{Synthetic ($L=13$)}\\
\cmidrule(lr){2-4}\cmidrule(lr){5-7}
Gates drawn from & median & min & Frac.\ $\le$ & median & min & Frac.\ $\le$\\
\midrule
All gene pairs & 0.145 & 0.035 & 0.00 & 0.930 & 0.531 & 0.10\\
Cross-cell pairs only & 0.142 & 0.019 & 0.01 & 0.676 & 0.517 & 0.40\\
Within-cell pairs only & 0.179 & 0.051 & 0.00 & 1.905 & 1.764 & 0.00\\
\bottomrule
\end{tabular}
\end{table}

On the ovarian data the selected four-gate topology beats every random four-gate topology drawn from the full pair set and 99\% of those drawn from cross-cell pairs, so the data-driven search singles out specific edges that the data favors over random topologies of the same size. On the synthetic data, 13 of the 16 possible cross-cell pairs are used, so random cross-cell topologies of that size are nearly the same set and the test has no power (hence the large Frac.\ $\le$). Section~\ref{si:synthetic} uses a single-edge ranking instead.

\section{Gate-Contribution Analysis}\label{si:contrib}
Let $\tau=(e_1,\dots,e_L)$, where $e_\ell=(c_\ell,t_\ell)$ is the $\ell$-th gate (a control$\to$target qubit pair, as in Section~\ref{si:model}), with angles $\theta$, let $\mathcal L_0=\mathcal L(\emptyset)$ and $\mathcal L_\tau=\mathcal L(\tau,\theta)$, and write $\Delta_{\rm tot}=\mathcal L_0-\mathcal L_\tau$.
\begin{definition}[Contribution measures]
\begin{align}
\text{Sequential (order-dependent):}\quad & s_\ell=\mathcal L(e_{1:\ell-1})-\mathcal L(e_{1:\ell}),\qquad \textstyle\sum_\ell s_\ell=\Delta_{\rm tot},\\
\text{Isolated:}\quad & i_\ell=\mathcal L_0-\mathcal L(e_\ell),\\
\text{Leave-one-out (gate removal):}\quad & r_\ell=\mathcal L(\tau\setminus e_\ell)-\mathcal L_\tau,
\end{align}
with the angles of the remaining gates held at their optimized values. Percentages are reported relative to $\Delta_{\rm tot}$, i.e.\ $100\,s_\ell/\Delta_{\rm tot}$ and $100\,r_\ell/\Delta_{\rm tot}$.
\end{definition}
The sequential shares sum to 100\% but depend on the order of the gates. The leave-one-out shares are order-independent (each is a proper gate-removal experiment in the model) but need not sum to 100\% because gates interact. None of these quantities is a biological intervention effect: they are counterfactuals \emph{within the fitted model}.

\subsection{Shot Noise}\label{si:shot}
Because marginalizing shot counts preserves multinomial statistics, each register's marginal distribution with $K$ patterns estimated from $N_s$ shots follows $\hat P\sim\mathrm{Mult}(N_s,P)/N_s$. A second-order expansion (Delta Method) gives, 
\begin{gather}
\mathbb E\bigl[\KL(\hat P\|Q)\bigr]\approx\KL(P\|Q)+\frac{K_+-1}{2N_s},\\
\mathrm{Var}\bigl[\KL(\hat P\|Q)\bigr]\approx\frac1{N_s}\Bigl[\sum_sP(s)\log^2\tfrac{P(s)}{Q(s)}-\KL(P\|Q)^2\Bigr],
\label{eq:si_shot}
\end{gather}
where $K_+$ is the number of patterns with $P(s)>0$, and the term $(K_+-1)/(2N_s)$ is the Miller-Madow bias correction. For the ovarian panel ($K=16$ and $8$, $N_s=5,000$), the combined bias across both registers is $\approx0.002$. The Monte-Carlo standard deviation (300 replicates) is 0.0073 at baseline and 0.0030 at the optimum. Contributions smaller than $\approx2\sigma$ ($\approx0.006$) cannot be resolved from 5,000 shots, which is why all SI gate-contribution analyses are evaluated using exact state-vector probabilities, with 5,000-shot sampled estimators reserved for quantifying shot noise and selected main-text figures.

\begin{figure}[h!]
\centering
\includegraphics[width=0.7\textwidth]{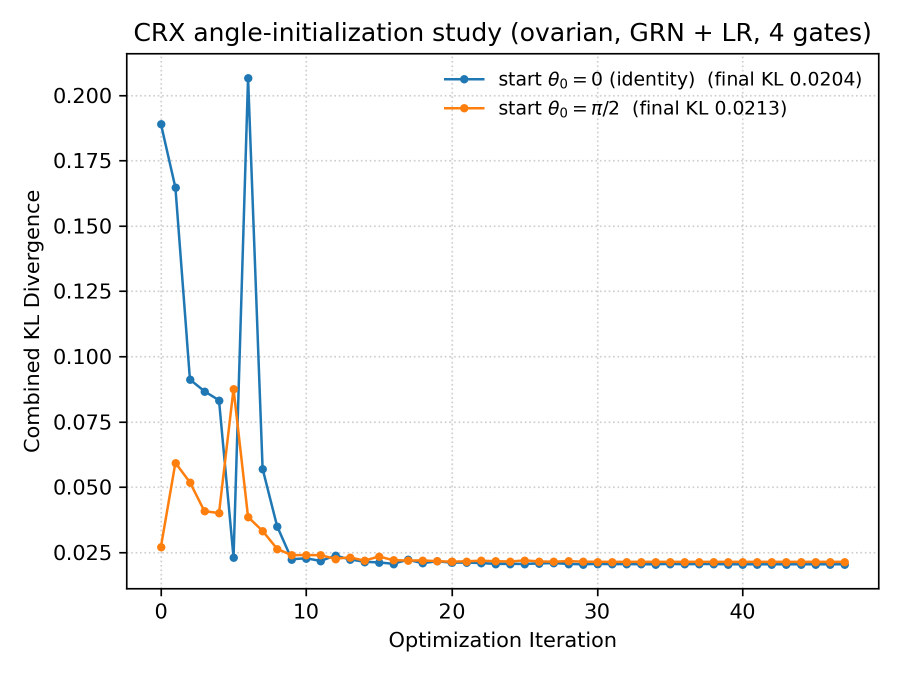}
\caption{\textbf{Angle-optimization convergence for the final L1 ovarian topology (4 gates, 5,000 shots, COBYLA) from $\theta=0$ and $\theta=\pi/2$.} Both initializations reach an equivalent loss (0.0204 and 0.0213), indicating no initialization-dependent local minimum for this topology.}\label{fig:si_init}
\end{figure}

\begin{figure}[h!]
\centering
\includegraphics[width=\textwidth]{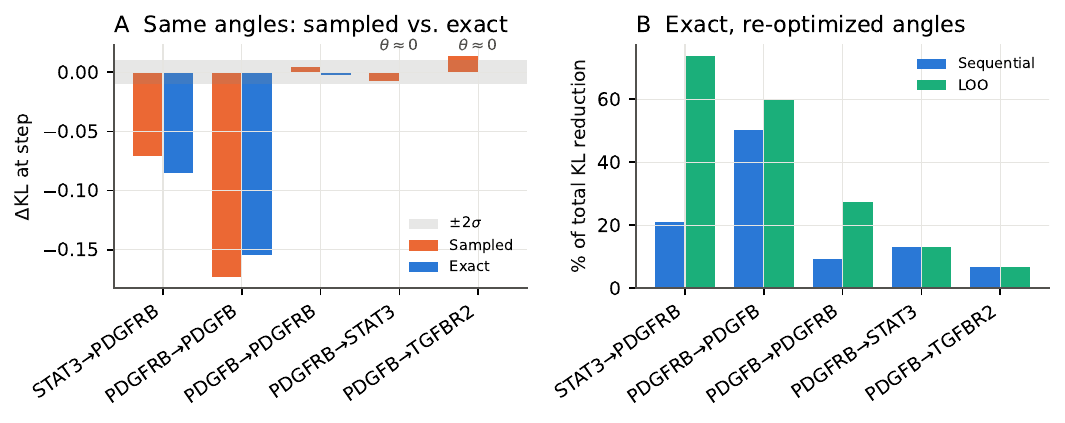}
\caption{\textbf{Exact versus sampled gate contributions.} (A) Step-wise KL divergence change ($\Delta\mathrm{KL}$) for the multi-epoch topology under L2 encoding, evaluated at the shot-sampled angles: 5,000-shot sampled estimates (orange) versus exact state-vector probabilities (blue) at identical angles. The shaded gray horizontal band denotes the $\pm2\sigma$ shot-noise threshold ($\approx0.006$), within which two gates frozen near identity ($\theta \approx 0$) produce non-zero artifacts under shot sampling. (B) Sequential (blue) and leave-one-out (LOO, aqua) gate contributions following exact angle re-optimization from $\theta = \pi/2$ (matching Table 3), expressed as a percentage of total KL reduction.}\label{fig:si_exact}
\end{figure}

\paragraph{Exact evaluation.} Under L2 encoding, evaluating the multi-epoch topology at the shot-sampled angles reveals that two gates sit near identity ($\theta\approx10^{-7}$ and $10^{-5}$~rad) and contribute zero under exact state-vector probabilities, confirming that their sampled non-zero drops were shot-noise artifacts (Fig.~\ref{fig:si_exact}A). Re-optimizing this topology directly on the exact objective from $\theta=\pi/2$ activates all gates and lowers the loss from 0.085 to 0.020 (Fig.~\ref{fig:si_exact}B). Identical optimization behavior holds under L1 encoding: optimizing from $\theta=0$ and from $\theta=\pi/2$ reaches equivalent loss minima across both encodings, ruling out initialization-dependent local traps (Fig.~\ref{fig:si_init}). Full per-gate exact and leave-one-out (LOO) contributions for all ovarian topologies across both L1 and L2 encodings are reported in Table~\ref{tab:si_contrib_l1}.

\begin{table}[htbp]
\centering\small
\caption{\textbf{Exact gate-contribution analysis for both amplitude encodings and both search algorithms (ovarian).} Per-gate share of the total KL reduction (\% Total), isolated reduction (Iso.\ Red.), and leave-one-out contribution (LOO \%), evaluated on exact state-vector probabilities with exactly optimized gate angles. The L2 values are those summarized in Table~3. Baselines differ by encoding (L1 0.194, L2 0.326), so absolute reductions are not comparable across encodings; the rankings and percentages are.}
\label{tab:si_contrib_l1}
\setlength{\tabcolsep}{5pt}
\begin{tabular}{lrrrr}
\toprule
Gate & $\theta$ (rad) & \% Total & Iso.\ Red. & LOO \% \\
\midrule
\multicolumn{5}{l}{\emph{L1 encoding, N-Wise} \quad $0.194\to0.0205$ (89.4\%)}\\
STAT3$\to$PDGFRB & 1.475 & 19.5 & 0.0338 & 26.4\\
PDGFRB$\to$PDGFB & 1.649 & 70.6 & 0.1054 & 57.8\\
PDGFB$\to$PDGFRB & 0.000 & 0.0 & 0.0000 & 0.0\\
TGFBR1$\to$PDGFB & 0.956 & 0.8 & 0.0267 & 1.6\\
TGFBR1$\to$TGFB1 & 1.246 & 5.0 & 0.0054 & 4.9\\
TGFB1$\to$TGFBR2 & 2.029 & 4.1 & 0.0069 & 4.1\\
\midrule
\multicolumn{5}{l}{\emph{L1 encoding, Multi-Epoch} \quad $0.194\to0.0212$ (89.0\%)}\\
STAT3$\to$PDGFRB & 1.486 & 19.6 & 0.0338 & 28.6\\
PDGFRB$\to$PDGFB & 1.802 & 71.7 & 0.1112 & 75.2\\
PDGFB$\to$TGFBR2 & 1.196 & 4.2 & 0.0038 & 4.2\\
TGFBR1$\to$TGFB1 & 1.399 & 4.5 & 0.0051 & 4.5\\
\midrule
\multicolumn{5}{l}{\emph{L2 encoding, N-Wise} \quad $0.326\to0.0298$ (90.9\%)}\\
PDGFB$\to$HES4 & 3.142 & $-2.5$ & $-0.0074$ & 14.4\\
PDGFB$\to$PDGFRB & 1.135 & 1.0 & 0.0031 & 1.4\\
PDGFRB$\to$PDGFB & 2.523 & 47.5 & 0.1346 & 60.3\\
PDGFB$\to$TGFBR2 & 1.909 & $-14.9$ & 0.0038 & 6.9\\
HES4$\to$PDGFRB & 1.583 & 50.7 & 0.0831 & 79.1\\
TGFBR2$\to$PDGFB & 0.000 & 0.0 & 0.0000 & 0.0\\
STAT3$\to$HES4 & 1.253 & 1.4 & $-0.0104$ & 12.1\\
PDGFRB$\to$STAT3 & 1.702 & 16.8 & 0.0319 & 16.8\\
\midrule
\multicolumn{5}{l}{\emph{L2 encoding, Multi-Epoch} \quad $0.326\to0.0198$ (93.9\%)}\\
STAT3$\to$PDGFRB & 1.603 & 20.9 & 0.0641 & 73.7\\
PDGFRB$\to$PDGFB & 1.294 & 50.1 & 0.0856 & 59.8\\
PDGFB$\to$PDGFRB & 2.327 & 9.3 & 0.0088 & 27.4\\
PDGFRB$\to$STAT3 & 1.921 & 13.0 & 0.0364 & 13.0\\
PDGFB$\to$TGFBR2 & 1.694 & 6.6 & 0.0032 & 6.6\\
\bottomrule
\end{tabular}
\end{table}

\subsection{Reproducibility: Bootstrap Stability of the Edges}\label{si:boot}
\begin{figure}[htbp]
\centering
\includegraphics[width=\textwidth]{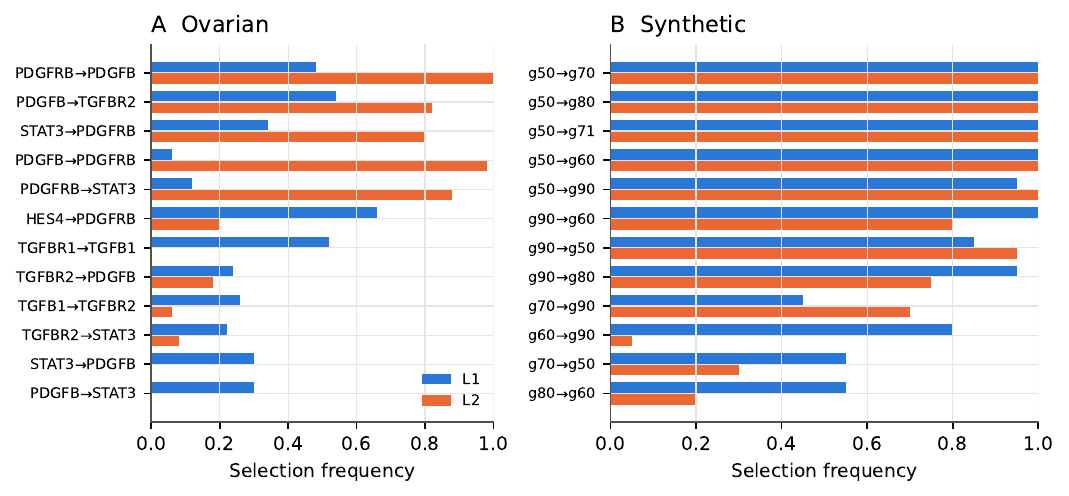}
\caption{\textbf{Bootstrap edge stability.} Edge selection frequency over cell-level resamples (full pipeline re-run per replicate, exact scoring) under L1 and L2 encodings. (A) Ovarian dataset (50 resamples); (B) Synthetic dataset (20 resamples).}\label{fig:si_boot}
\end{figure}

To separate stable edges from search artifacts, the complete pipeline was executed on 50 ovarian and 20 synthetic bootstrap replicates (resampled with replacement within cell groups). Selection frequencies separate per-edge stability from higher-level structural invariants (Fig.~\ref{fig:si_boot}, Table~\ref{tab:si_boot}).

\paragraph{Structural stability vs. edge identification.} In every ovarian replicate under both encodings, at least one cross-cell edge targets \texttt{PDGFRB} in fibroblasts, gated by \texttt{STAT3} or \texttt{HES4} (Table~\ref{tab:si_boot}). Under L1 encoding, \texttt{STAT3} and \texttt{HES4} have similar basal activities ($\pi=0.68$ and $0.60$) and substitute for one another across resamples (\texttt{STAT3}$\to$\texttt{PDGFRB} 34\%, \texttt{HES4}$\to$\texttt{PDGFRB} 66\%). L2's squared weighting separates these activities more strongly, concentrating selection onto individual edges at the expense of fitting the escort distribution. On synthetic data, core sender edges ($g_{50}\to g_{60}, g_{70}, g_{71}, g_{80}$) achieve 100\% stability under both encodings (Fig.~\ref{fig:si_boot}b).

\begin{table}[htbp]
\centering\small
\caption{\textbf{Structural bootstrap invariants on ovarian data (50 resamples).} Values represent selection frequencies (0.00--1.00) for L1 and L2 amplitude encodings; the bottom row reports mean gate count [range].}\label{tab:si_boot}
\begin{tabular}{lrr}
\toprule
Structural Event / Metric & L1 & L2\\
\midrule
Any cross-cell edge from cancer cells to \texttt{PDGFRB} (\texttt{STAT3} or \texttt{HES4}) & 1.00 & 1.00\\\texttt{PDGFB} targeted across cell types & 0.70 & 1.00\\
\texttt{TGFBR2} targeted across cell types & 0.76 & 0.86\\
$\ge2$ edges within \{\texttt{PDGFB}, \texttt{PDGFRB}, \texttt{STAT3}\} & 0.54 & 1.00\\
Any edge involving a TGF-$\beta$ pathway gene & 0.90 & 0.88\\
Mean gate count [range] & 4.5 [2--10] & 6.0 [4--10]\\
\bottomrule
\end{tabular}
\end{table}

\section{Benchmark by Biological Scope: GRN, LR and GRN+LR}\label{si:scopes}
To evaluate the relative contributions of within-cell and cross-cell signaling, full search and angle optimization were executed separately across three biological scopes: GRN+LR (all pairs, density-screened), GRN only (within-cell pairs, optimized per cell type), and LR only (all cross-cell pairs in both directions). For each final topology, three complementary contribution measures were computed: cumulative reduction as gates are added sequentially, isolated reduction of each gate alone, and leave-one-out (LOO) loss increase when the gate is removed from the complete circuit. 

Table~\ref{tab:si_scopes} reports results on ovarian data recomputed with exact probabilities. Across every scope permitting cross-cell interactions, the \texttt{PDGFRB}$\to$\texttt{PDGFB} edge represents the primary contributor (62--75\% of total reduction upon LOO removal), followed by a cancer-cell control gene (\texttt{STAT3} or \texttt{HES4}) targeting \texttt{PDGFRB}. Within-cell gates alone reach a 71\% reduction via the \texttt{STAT3}$\leftrightarrow$\texttt{PDGFB} feedback pair.

\begin{table}[htbp]
\centering\small
\caption{\textbf{Benchmark A on the ovarian data (L1, multi-epoch, exact probabilities).} ``Cum.\ \%'': cumulative reduction after this gate as a share of the initial KL (0.194). ``Step \%'': this gate's step as a share of the total reduction. ``Iso.'': reduction by the gate alone. ``LOO \%'': loss increase when the gate is removed, as a share of the total reduction.}\label{tab:si_scopes}
\setlength{\tabcolsep}{3pt}\begin{tabular}{p{2.6cm}lrrrrr}
\toprule
Scope (final KL, \% red.) & Gate & $\theta$ & Cum.\ \% & Step \% & Iso. & LOO \%\\
\midrule
GRN+LR (0.021, 89.0\%) & STAT3$\to$PDGFRB & 1.486 & 17.4 & 19.6 & 0.034 & 28.6\\
 & PDGFRB$\to$PDGFB & 1.802 & 81.3 & 71.7 & 0.111 & 75.2\\
 & PDGFB$\to$TGFBR2 & 1.196 & 85.1 & 4.2 & 0.004 & 4.2\\
 & TGFBR1$\to$TGFB1 & 1.399 & 89.0 & 4.5 & 0.005 & 4.5\\
\midrule
GRN only (0.056, 71.1\%) & STAT3$\to$PDGFB & 1.325 & 45.5 & 64.0 & 0.088 & 104.6\\
 & PDGFB$\to$STAT3 & 1.625 & 71.1 & 36.0 & $-0.006$ & 36.0\\
\midrule
LR only (0.015, 92.1\%) & HES4$\to$PDGFRB & 1.531 & 17.4 & 18.9 & 0.034 & 24.2\\
 & PDGFRB$\to$PDGFB & 1.733 & 78.9 & 66.8 & 0.109 & 62.4\\
 & TGFBR2$\to$STAT3 & 1.995 & 83.4 & 4.9 & 0.019 & 6.2\\
 & TGFB1$\to$TGFBR2 & 1.966 & 88.1 & 5.1 & 0.007 & 5.5\\
 & TGFBR1$\to$TGFB1 & 1.131 & 92.1 & 4.3 & 0.005 & 4.3\\
\bottomrule
\end{tabular}
\end{table}

\section{Classical and Tensor-Network Baselines: Definitions}\label{si:baselines}
To systematically evaluate model performance against classical and tensor-network baselines, models are benchmarked across two distinct capacity regimes: independent site transformations ($D=1$, testing the null hypothesis that single-gene self-regulation alone drives the co-culture shift) and minimal correlated networks ($D=2$, testing generic low-rank cross-gene interactions). All baselines model each cell-type register independently and are scored using the exact per-register KL divergence~\eqref{eq:si_loss}, summed across both registers. For a register of $n$ genes, let $p = \hat{p}^{\mathrm{mo}}$ denote the basal monoculture probability, $q = \hat{p}^{\mathrm{co}}$ the target co-culture probability, and $v_0 = \sqrt{p}$ the baseline L1 amplitude vector.

\begin{definition}[Entropic Optimal Transport, OT]
Using the normalized Hamming distance matrix $M_{xy} = d_H(x,y)/n$ between binary state patterns and entropic regularization $\varepsilon_{\mathrm{OT}} = 0.05$, the optimal coupling matrix $\Gamma^\star$ is solved via Sinkhorn iterations~\cite{cuturi2013sinkhorn,flamary2021pot}:
\begin{equation}
\Gamma^\star = \operatorname*{argmin}_{\Gamma \in \Pi(p,q)} \langle \Gamma, M \rangle - \varepsilon_{\mathrm{OT}} H(\Gamma).
\end{equation}
The predicted state distribution $\hat{q}$ is formed by the barycentric push-forward $T(x) = \sum_y \Gamma^\star_{xy} y / \sum_y \Gamma^\star_{xy}$, rounded coordinate-wise to $\{0,1\}^n$, accumulating probability as $\hat{q}(\operatorname{round}(T(x))) \mathrel{+}= p(x)$. Capped variants restrict the coupling matrix to the top-$k$ largest entries of $\Gamma^\star$ (un-coupled source patterns remain invariant). OT is conditioned on the basal state but returns no directed gene interaction network.
\end{definition}

\begin{definition}[Tensor-Network Transform, TN]
A real Matrix Product Operator (MPO) $W$ of bond dimension $D$~\cite{verstraete2004mpdo} acts on the basal amplitude vector:
\begin{equation}
\psi = \frac{W v_0}{\|W v_0\|_2}, \quad \hat{q} = \psi^{\odot 2}.
\end{equation}
At each site $j$, $W$ is composed of a 4-index local tensor $A_j \in \mathbb{R}^{D \times D \times 2 \times 2}$, indexed as $(A_j)_{\alpha_{j-1}, \alpha_j, s'_j, s_j}$ with virtual bond indices $\alpha_{j-1}, \alpha_j \in \{1, \dots, D\}$ and physical gene state indices $s'_j, s_j \in \{0, 1\}$.
\begin{itemize}
    \item At $D=1$, the virtual bond dimensions collapse ($1 \times 1$), so $A_j \in \mathbb{R}^{2\times 2}$ reduces to a standard single-gene $2\times 2$ matrix $W = \bigotimes_{j=1}^n A_j$, modeling independent per-gene re-tuning with zero cross-gene correlations.
    \item At $D=2$, $A_j \in \mathbb{R}^{2\times 2\times 2\times 2}$ retains a 2-dimensional virtual bond between adjacent sites, allowing $W$ to introduce minimal low-rank entangling interactions between genes.
\end{itemize}
The model is non-unitary and conditioned on the basal state.
\end{definition}

\begin{definition}[Born Machine]
A real Matrix Product State (MPS) $\phi$ with bond dimension $D$ is fitted directly to the target distribution $q$, yielding $\hat{q} = \phi^{\odot 2} / \|\phi\|_2^2$~\cite{han2018mps}. At each site $j$, $\phi$ is composed of a 3-index local tensor $\phi_j \in \mathbb{R}^{D \times D \times 2}$, indexed as $(\phi_j)_{\alpha_{j-1}, \alpha_j, s_j}$ with virtual bond indices $\alpha_{j-1}, \alpha_j \in \{1, \dots, D\}$ and physical gene state index $s_j \in \{0, 1\}$.
\begin{itemize}
    \item At $D=1$, the virtual bond dimensions collapse ($1 \times 1$), so $\phi_j \in \mathbb{R}^2$ reduces to a 2-element single-site vector $(u_j, v_j)^\top$. Normalization yields an independent product distribution $\hat{q}(x) = \prod_{j=1}^n \beta_j^{x_j}(1-\beta_j)^{1-x_j}$ with parameter $\beta_j = v_j^2 / (u_j^2 + v_j^2) \in [0,1]$ per site.
    \item At $D=2$, $\phi_j \in \mathbb{R}^{2 \times 2 \times 2}$ retains a 2-dimensional virtual bond between adjacent sites, allowing $\phi$ to capture low-rank multi-gene correlations.
\end{itemize}
This baseline is unconditional with respect to $v_0$ and serves as a descriptive reference for direct target-fitting.
\end{definition}

\begin{definition}[Local-$R_Y$ Transform]
The state transformation is defined as $\psi = \bigl(\bigotimes_{j=1}^n R_Y(\theta_j)\bigr) v_0$. This corresponds to the unitary, $D=1$ subset of the TN-transform family, modeling independent single-gene rotation (self-regulation) with zero interaction network topology ($L=0$).
\end{definition}

\paragraph{Optimization.} Gate rotation angles in QuantumXCT are optimized using COBYLA initialized from the topology-search start ($\theta = \pi/2$). Parameters for Tensor-Network ($D=1,2$) and Born-Machine ($D=1,2$) baselines are optimized on the exact KL divergence objective using L-BFGS-B from seeded initializations. The Local-$R_Y$ baseline is optimized using L-BFGS-B with 20 random restarts plus a zero-rotation start. Optimal Transport uses fixed entropic regularization $\varepsilon_{\mathrm{OT}} = 0.05$ with Sinkhorn fixed-point iterations and no trainable parameters.

\section{Comparison with Classical and Tensor-Network Baselines}\label{si:budget}
We compare QuantumXCT against the baselines defined in Section~\ref{si:baselines} using the exact per-register loss across comparable capacity regimes. For tensor-network baselines, virtual bond dimension $D$ determines model capacity: $D{=}1$ represents a factorized product state modeling independent single-gene self-regulation, while $D{=}2$ is the minimal entangled capacity capturing low-rank cross-gene correlations. We also include the local self-regulation model (one free $R_Y$ rotation per gene, zero interaction topology) as the unitary $D{=}1$ reference, alongside entropic optimal transport ($m{=}1$, capped at QuantumXCT's gate budget $L$). The Born machine is listed as an unconditioned oracle reference because it is fitted directly to target co-culture marginals rather than conditioned on the basal state $v_0$. Only QuantumXCT returns a sparse, directed, and biologically interpretable interaction network.

Table~\ref{tab:si_capacity} reports exact-scoring losses and parameter counts across both datasets. QuantumXCT achieves KL divergence loss comparable to low-capacity product models while utilizing the fewest parameters, whereas higher-capacity $D{=}2$ tensor networks lower the loss only by using substantially more parameters. Crucially, QuantumXCT is the only evaluated method that yields a biologically interpretable, directed interaction network.

\begin{table}[htbp]
\centering
\caption{\textbf{Method comparison at matched capacity (exact scoring).} For tensor-network baselines, $D$ denotes the virtual bond dimension: $D{=}1$ represents a factorized product state (independent per-gene re-tuning) and $D{=}2$ introduces minimal low-rank cross-gene entangling interactions. For Optimal Transport, $m=1$ caps the transport plan to $1 \times L$ couplings per register, matching QuantumXCT's gate budget $L$. ``Params'' reports stored parameters/plan entries per register ($n_{\mathrm{qubits}} = 7$ for Ovarian, $n_{\mathrm{qubits}} = 6$ for Synthetic for Local $R_Y$); ``Cond.''~= conditioned on the basal state; ``Net.''~= returns a directed interaction network. KL is the summed per-register $D_{\mathrm{KL}}(P_{\mathrm{model}}\|Q_{\mathrm{co}})$ evaluated on exact output probability vectors.}
\label{tab:si_capacity}
\begin{tabular}{lrrcc}
\toprule
Method & KL (c1+c2) & Params & Cond. & Net.\\
\midrule
\multicolumn{5}{l}{\textit{Ovarian cancer--fibroblast}}\\
    \textbf{QuantumXCT (best effort)} & 0.0212 & 4 & yes & yes\\
    Local self-regulation ($D{=}1$) & 0.0066 & 7 & yes & no\\
    TN transform, $D{=}1$ (product) & 0.0007 & 28 & yes & no\\
    TN transform, $D{=}2$ (pairwise) & $<10^{-4}$ & 80 & yes & no\\
    Born machine, $D{=}1$ (product) & 0.0194 & 14 & no$^{*}$ & no\\
    Born machine, $D{=}2$ (pairwise) & 0.0018 & 40 & no$^{*}$ & no\\
    Optimal transport ($m{=}1$) & 0.1939 & 8 & yes & no\\
\midrule
\multicolumn{5}{l}{\textit{Synthetic benchmark}}\\
    \textbf{QuantumXCT (best effort)} & 0.5324 & 12 & yes & yes\\
    Local self-regulation ($D{=}1$) & 0.0071 & 6 & yes & no\\
    TN transform, $D{=}1$ (product) & 0.0021 & 24 & yes & no\\
    TN transform, $D{=}2$ (pairwise) & $<10^{-4}$ & 64 & yes & no\\
    Born machine, $D{=}1$ (product) & 0.0073 & 12 & no$^{*}$ & no\\
    Born machine, $D{=}2$ (pairwise) & 0.0032 & 32 & no$^{*}$ & no\\
    Optimal transport ($m{=}1$) & 1.8823 & 24 & yes & no\\
\bottomrule
\end{tabular}
\par\smallskip
{\footnotesize $^{*}$The Born machine is fitted directly to target co-culture marginals without conditioning on basal state $v_0$, serving as an oracle descriptive reference.}
\end{table}

\section{Synthetic Benchmark: Edge and Direction Recovery}\label{si:synthetic}
\paragraph{Setup and metrics.} The synthetic 6-gene panel $\{g_{50},g_{90}\mid g_{60},g_{70},g_{71},g_{80}\}$ contains 5 direct programmed edges ($50\to60, 60\to70, 70\to71, 70\to80, 80\to90$) and 13 transitive causal paths among 30 ordered pairs. Predicted edge sets $E$ are evaluated against ground truth $G$ via precision $|E\cap G|/|E|$, recall $|E\cap G|/|G|$, direction accuracy $|E\cap G|/|\{e\in E:e\in G\text{ or }\bar e\in G\}|$, and hypergeometric significance $p_{\mathrm{hyp}}$. Pairwise edge ranking uses the single-edge loss reduction $i^\star_{c\to t}=\mathcal L_0-\min_\theta\mathcal L((c,t),\theta)$ evaluated by ROC-AUC.

\begin{table}[htbp]
\centering\small
\caption{\textbf{Edge recovery on the synthetic benchmark.}}\label{tab:si_recovery}
\setlength{\tabcolsep}{4pt}\begin{tabular}{lrrrrrr}
\toprule
 & \multicolumn{3}{c}{Direct edges ($|G|=5$)} & \multicolumn{3}{c}{Causal paths ($|G|=13$)}\\
Topology & precision & recall & $p_{\rm hyp}$ & precision & dir.\ acc. & $p_{\rm hyp}$\\
\midrule
Exact L1, all scope (12 gates) & 2/12 & 0.40 & 0.68 & 7/12 & 0.58 & 0.16\\
Exact L1, cross-cell scope (10 gates) & 2/10 & 0.40 & 0.55 & 6/10 & 0.67 & 0.18\\
Exact L2, all scope (12 gates) & 1/12 & 0.20 & 0.94 & 7/12 & 0.58 & 0.16\\
Exact L2, cross-cell scope (8 gates) & 2/8 & 0.40 & 0.40 & 6/8 & 0.75 & 0.045\\
\bottomrule
\end{tabular}
\end{table}

\begin{figure}[htbp]
\centering
\includegraphics[width=0.85\textwidth]{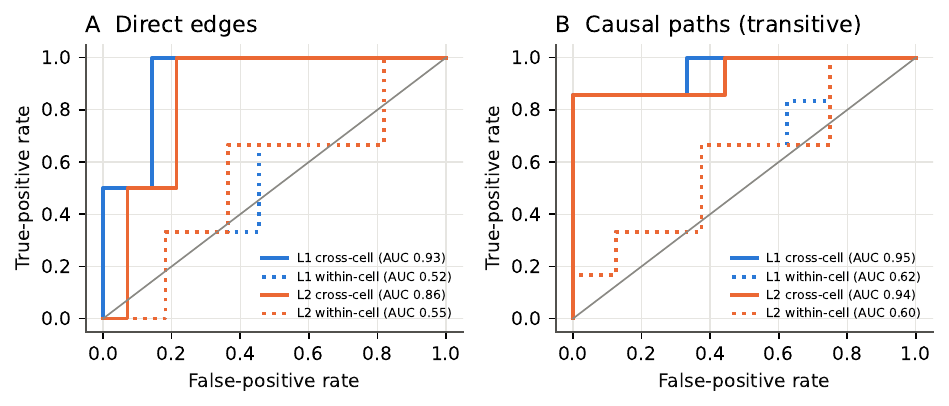}
\caption{\textbf{ROC of single-edge scores on the synthetic benchmark.} ROC curves evaluating cross-cell (solid lines) and within-cell (dotted lines) gene pairs under $L_1$ (blue) and $L_2$ (orange) normalization against (A) direct edges ($|G|=5$) and (B) transitive causal paths ($|G|=13$).}\label{fig:si_roc}
\end{figure}

Multi-gate topologies reliably recover causal structure over chance (Table~\ref{tab:si_recovery}; 6--7 of 8--12 edges lie on true causal paths, with $p = 0.045$ for cross-cell $L_2$), frequently introducing expected ``shortcut'' edges from sender $g_{50}$ to downstream targets (Section~\ref{si:ident}). Single-edge scores accurately identify intercellular interactions (Figure~\ref{fig:si_roc}; top-ranked edge $50\to60$, cross-cell AUCs 0.86--0.95), whereas intracellular GRN edges remain unidentifiable from marginals (AUCs 0.52--0.62). QuantumXCT results are therefore best understood as recovering intercellular causal structure rather than full internal wiring diagrams.

\section{Comparison with CCC Tools and Experimental Evidence}\label{si:ccc}
\paragraph{Established tools on the same data.} We evaluated LIANA+~\cite{dimitrov2024liana} (consensus rank aggregation with differential contrast) and scTenifoldXct~\cite{yang2023sctenifoldxct} (manifold alignment of mono- and co-culture GRNs) alongside CellChat~\cite{jin2021inference,jin2025cellchat} under default parameters on the ovarian dataset. scTenifoldXct ranks \texttt{PDGFB-PDGFRB} 9th among 49 significant differential interactions ($q=2.9\times10^{-6}$, manifold distance $\mathrm{diff}^2 = 1.15\times10^{-5}$). In LIANA+, the cancer$\to$fibroblast pair \texttt{PDGFB-PDGFRA\_PDGFRB} increases in co-culture ($\Delta$ strength $=4.97$), and \texttt{TGFB1-TGFBR1\_TGFBR2} is among the top-ranked increases ($\Delta=3.29$), consistent with QuantumXCT's secondary TGF-$\beta$ edge (\texttt{TGFBR1}$\to$\texttt{TGFB1}). CellChat infers communication probabilities per condition using law-of-mass-action functions under default thresholds ($\mathrm{thresh.fc} = 0.05$). While CellChat strongly confirms the \texttt{PDGFB-PDGFRB} axis ($prob = 0.0012$, $p < 10^{-17}$), its mass-action complex scoring misses co-culture TGF-$\beta$ signaling: because heteromeric complex rules require joint expression across all subunits, a minor negative log-fold change in the \texttt{ACVR1B} subunit ($-0.092$) inclines the interaction exclusively toward mono-culture ($prob = 6.5\times10^{-4}$ in mono-culture; zero called in co-culture; Table~\ref{tab:si_ccc}, footnote $a$). As previously demonstrated, classical mass-action scoring routinely mistakes coincidental marginal co-expression for genuine interaction shift~\cite{romero2026qsimcells}. By evaluating continuous differential statistics across individual subunits rather than requiring joint mass-action thresholds, LIANA+ and QuantumXCT robustly recover both the PDGF and TGF-$\beta$ signaling channels ($\Delta = 3.29$--$3.92$), directly aligning with independent perturbation experiments.

\begin{table}[htbp]
\centering\small
\caption{\textbf{Cross-method comparison of the two interaction axes on the ovarian data.} Each method is reported in its own metric: CellChat communication probability (co-culture); LIANA+ differential interaction strength ($Co-Mo$); scTenifoldXct manifold-distance $\mathrm{diff}^2$ (significant differential pairs); QuantumXCT exact leave-one-out KL contribution (L1 encoding). ``--'' indicates the pair is not reported by that method.}\label{tab:si_ccc}
\begin{tabular}{lcccc}
\toprule
Ligand--receptor & CellChat (prob) & LIANA+ ($Co-Mo$) & scTenifoldXct ($\mathrm{diff}^2$) & QuantumXCT (KL)\\
\midrule
PDGFB--PDGFRB & 0.0012 & 4.97 & $1.15\times10^{-5}$ & 0.130\\
TGFB1--TGFBR1 & -- & 3.29 & -- & 0.005\\
TGFB1--TGFBR2 & --$^{a}$ & 3.92 & -- & --\\
\midrule
\multicolumn{5}{p{11.5cm}}{\footnotesize $^{a}$CellChat assigns \texttt{TGFB1}$\to$\texttt{(ACVR1B+TGFBR2)} exclusively to mono-culture ($prob = 6.5\times 10^{-4}$, fibroblast$\to$cancer) and misses co-culture induction because a receptor subunit logFC ($-0.092$) fails its strict \texttt{AND} complex threshold ($\mathrm{thresh.fc} = 0.05$); no co-culture \texttt{TGFB1} pair is called.}\\
\bottomrule
\end{tabular}
\end{table}

\paragraph{Experimental and perturbation evidence.} Independent experiments on these co-culture data by Mori et al.~\cite{mori2024targeting} directly validate our inferred network: (i) co-culture induces activating PDGFR$\beta$ phosphorylation in CAFs (phenocopied by purified PDGF-BB); (ii) CRISPR knockout of \emph{PDGFRB} in CAFs abolishes proliferation and the myCAF marker FAP-$\alpha$; (iii) CAF \emph{PDGFRB} knockout \emph{reduces PDGFB expression in cancer cells}, establishing a positive feedback loop; and (iv) co-culture strongly induces CAF TGF-$\beta$ signaling. The primary QuantumXCT edge, \texttt{PDGFRB}(fibroblast)$\to$\texttt{PDGFB}(cancer), directly mirrors (iii): cancer-cell \emph{PDGFB} shifts as a response gated by fibroblast \emph{PDGFRB} activity (Section~\ref{si:ident}). Downstream, PDGFR$\beta$ signals via STAT3~\cite{heldin2013pdgfr}, a key effector in CAF activation~\cite{sahai2020caf}, supporting the \texttt{STAT3}--\texttt{PDGFRB} link. These independent findings provide orthogonal consistency checks, though QuantumXCT itself generates data-driven hypotheses rather than establishing formal causation (assumption A3).

\section{Scalability}\label{si:scal}
\begin{figure}[h]
\centering
\includegraphics[width=0.9\textwidth]{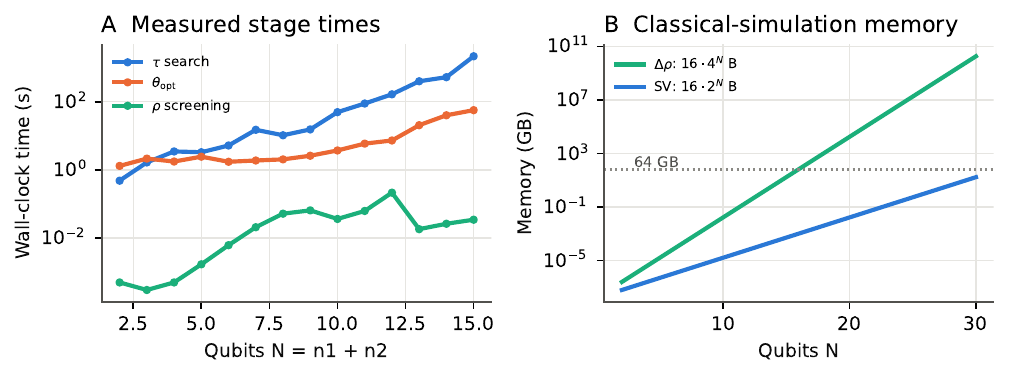}
\caption{\textbf{Computational scaling of the pipeline.} \textbf{(A)} Measured wall-clock time of each stage versus qubit number $N=n_1+n_2$ (synthetic data): topology search ($\tau$, blue), gate-angle optimization ($\theta_{\text{opt}}$, orange) and density-matrix screening ($\rho$, green). \textbf{(B)} Classical-simulation memory versus $N$: the dense density difference $\Delta\rho$, $16\cdot4^N$ bytes (green), and the statevector (SV), $16\cdot2^N$ bytes (blue); the dotted line marks 64~GB.}\label{fig:si_scaling}
\end{figure}

Let $N=n_1+n_2$ be the number of qubits (one per gene), $s_k\le\min(2^{n_k},\#\text{cells})$ the number of observed patterns, $|\mathcal C|\le N(N-1)$ the number of candidate gates, $L$ the circuit length and $E$ the number of epochs.
\begin{itemize}
\item \emph{Encoding.} Preparing an arbitrary $n$-qubit amplitude state needs $O(2^n)$ CNOTs~\cite{shende2006synthesis}; sparse states with $s$ non-zero amplitudes need $O(sn)$~\cite{gleinig2021sparse}. Because $s\le\#$cells, state preparation is bounded by data sparsity, but on hardware it dominates circuit depth (Section~\ref{si:hw}).
\item \emph{Screening.} The dense $\Delta\rho$ needs $16\cdot4^N$ bytes (about 1~GB at $N=13$, 68~GB at $N=16$); the sparse implementation needs $O((s_1s_2)^2)$, set by the data rather than $4^N$, and returns identical candidates. The threshold $\delta\rho$ is a tractability--resolution knob (main-text Appendix~B): it prunes little on the small panels here but controls the candidate-set size on larger ones.
\item \emph{Topology search.} Multi-epoch search evaluates $O(E\,|\mathcal C|\,L)$ circuits, i.e.\ $O(N^4)$ for $E\sim|\mathcal C|$ and $L\ll|\mathcal C|$. N-wise search adds $O(|\mathcal C|^n)$ per addition step, and the QUBO route needs $|\mathcal C|$ qubits for the VQE ansatz (42 at $\delta\rho=0.005$, which is why that row was infeasible; 12 at $\delta\rho=0.02$, where VQE-hybrid selected \texttt{STAT3}$\to$\texttt{PDGFRB}, \texttt{PDGFRB}$\to$\texttt{PDGFB}, 81.3\% reduction).
\item \emph{Cost per evaluation.} Exact classical evaluation costs $O(L\,2^N)$ time and $16\cdot2^N$ bytes (1~GB at $N=26$). On hardware, each evaluation costs $N_s$ shots at depth $O(2^{n_1}+2^{n_2}+L)$.
\item \emph{Sampling noise.} By~\eqref{eq:si_shot}, resolving a loss difference $\delta$ at $z$ standard errors needs $N_s\gtrsim z^2\,\mathrm{Var}_1/\delta^2$ with $\mathrm{Var}_1=\sum_sP\log^2(P/Q)-\KL^2$. For the ovarian panel, $\delta=0.005$ at $z=2$ needs $N_s\approx2\times10^4$. This $\delta^{-2}$ floor is why per-gate contributions, differences at the $0.005$ scale, are scored exactly, whereas the 5,000-shot estimator is reserved for the much larger loss reductions at the $0.05$--$0.17$ scale (Section~\ref{si:shot}).
\end{itemize}
Measured wall-clock times (Fig.~\ref{fig:si_scaling}) grow from seconds at $N=4$ to $\sim$45~min at $N=12$ with a deliberately short search (2 epochs, depth 4, 1000 shots), dominated by the topology search. Practical application to thousands of genes therefore requires (i) biological pre-selection of modules of $\lesssim10$--15 genes per cell type, and (ii) tiling many such modules. Each module is an independent QuantumXCT problem, so the cost is linear in the number of modules.

\paragraph{Trainability.} At the register sizes used here (the 7-qubit ovarian and 6-qubit synthetic panels, both $N\le10$), angle optimization is well behaved. COBYLA reaches an equivalent loss from both $\theta=0$ and $\theta=\pi/2$ initializations, with no evidence of an initialization-dependent local minimum (Fig.~\ref{fig:si_init}). Barren plateaus (exponentially vanishing gradients that impede training of deep, highly expressive circuits~\cite{mcclean2018barren,cerezo2021cost}) are not expected in this regime. The ansatz is shallow (a handful of CRX gates) and deliberately under-parameterized, and the two-stage strategy fixes the topology before the few continuous angles are optimized. For much larger panels the trainability of expressive ansatze becomes a genuine concern, requiring problem-informed initialization or local cost functions~\cite{cerezo2021cost}; the sparse, interpretable circuits favored by QuantumXCT mitigate this by construction.

\section{Post-Optimization Hardware Validation}\label{si:hw}
\paragraph{Protocol and noisy simulation.} The optimized ovarian circuit (L1, 4 gates) and baseline circuit (no gates) were evaluated using \texttt{StatePreparation} (avoiding reset operations) with optimization level 3 transpilation, dynamical decoupling (DD, XX sequence), gate/measurement Pauli twirling, and tensored readout error correction $\hat P=\Pi_\Delta\bigl[(\bigotimes_j A_j^{-1})P_{\rm raw}\bigr]$. On a 127-qubit noisy simulator (\texttt{FakeBrisbane}, 8,000 shots, transpiled depth 104 with 43 two-qubit gates dominated by state preparation), loss dropped from 0.148 to 0.035 after readout correction (76.4\% reduction; raw $0.125\to0.041$, 67.4\%), closely tracking the exact noiseless 87.8\% reduction ($0.194\to0.024$).

\paragraph{Hardware performance and mitigation sweep.} Executing the same circuits on the 127-qubit \texttt{ibm\_kingston} device (8,000 shots; transpiled depth 101, 43 two-qubit gates; physical qubit median $T_1=261\,\mu\text{s}$, $T_2=243\,\mu\text{s}$, single-qubit error $3.1\times 10^{-4}$, CZ error $2.1\times 10^{-3}$, readout error $0.6\%$) across five mitigation settings in a single calibration window demonstrates that unmitigated (raw) execution achieves an 89.3\% loss reduction ($0.162\to0.017$; Table~\ref{tab:hw_mit}), matching noiseless theory within shot noise. Because baseline readout error is low ($0.6\%$), tensored correction adds estimator variance without net benefit (84.7\% reduction), while DD and twirling insert pulse overhead that slightly degrades performance (72.8\% and 73.0\%; combined 64.9\%). The sparse 4-gate topology keeps execution well within hardware coherence budgets without requiring active error mitigation.

\begin{table}[htbp]
\centering
\caption{\textbf{Error-mitigation sweep on \texttt{ibm\_kingston} (optimized ovarian circuit, L1, 4 gates; 8,000 shots).} Loss reductions are relative to each setting's own baseline (no-gate) circuit. The noiseless reference is $0.194\to0.024$ (87.8\% reduction).}
\label{tab:hw_mit}
\small
\begin{tabular}{lrrr}
\toprule
Mitigation & Baseline KL & Optimized KL & Reduction \\
\midrule
Raw (none)              & 0.162 & 0.017 & 89.3\% \\
Readout correction      & 0.162 & 0.025 & 84.7\% \\
Dynamical decoupling    & 0.136 & 0.037 & 72.8\% \\
Pauli twirling          & 0.130 & 0.035 & 73.0\% \\
All combined            & 0.124 & 0.044 & 64.9\% \\
\midrule
Noiseless (exact)       & 0.194 & 0.024 & 87.8\% \\
\bottomrule
\end{tabular}
\end{table}

\paragraph{Zero-noise extrapolation (ZNE).} Global unitary folding $U\mapsto U(U^\dagger U)^n$ at scales $\lambda=1,3,5$ (with DD and twirling) yields monotonic loss growth ($0.049\to0.150\to0.240$; Table~\ref{tab:hw_zne}). Linear extrapolation to $\lambda\to0$ yields a residual loss of $0.003$ (quadratic $\approx0$), confirming that hardware residual is noise-limited rather than a structural model misspecification.

\begin{table}[htbp]
\centering
\caption{\textbf{Zero-noise extrapolation on \texttt{ibm\_kingston} (optimized circuit, DD and twirling on).} Noise scale is $\lambda=2n+1$ under global folding $U(U^\dagger U)^n$.}
\label{tab:hw_zne}
\small
\begin{tabular}{rrr}
\toprule
Noise scale $\lambda$ & Transpiled depth & Optimized KL \\
\midrule
1 & 101 & 0.049 \\
3 & 302 & 0.150 \\
5 & 508 & 0.240 \\
\midrule
$\to0$ (linear)    & -- & 0.003 \\
$\to0$ (quadratic) & -- & $\approx0$ \\
\bottomrule
\end{tabular}
\end{table}

\end{document}